\documentclass[iop,revtex4]{emulateapj}

\usepackage{amsmath}

\newcommand{\mytilde}{\raise.19ex\hbox{$\scriptstyle\sim$}}

\newcommand{\idcs}{IDCS~J1426+3508}
\newcommand{\spt}{SPT-CL~J2040--4451}
\shorttitle{SeeChange Weak-lensing}
\shortauthors{Jee et al.}

\begin{document}

\title{FIRST WEAK-LENSING RESULTS FROM ``SEE CHANGE": QUANTIFYING DARK MATTER IN THE TWO $Z\gtrsim1.5$ HIGH-REDSHIFT GALAXY CLUSTERS SPT-CL~J2040--4451 AND IDCS~J1426+3508} 

\author{M. JAMES JEE\altaffilmark{1,2}, JONGWAN KO\altaffilmark{3,4}, SAUL PERLMUTTER\altaffilmark{5} ANTHONY GONZALEZ\altaffilmark{6}, MARK BRODWIN\altaffilmark{7},\\ ERIC LINDER\altaffilmark{5}, AND PETER EISENHARDT\altaffilmark{8}}

\begin{abstract}
We present a weak-lensing study of SPT-CL~J2040--4451 and IDCS~J1426+3508 at $z=1.48$ and $1.75$, respectively. The two clusters were observed in our ``See Change" program, a {\it Hubble Space Telescope (HST)} survey of 12 massive high-redshift clusters aimed at high-$z$ supernova measurements and weak-lensing estimation of accurate cluster masses.
We detect weak but significant galaxy shape distortions using infrared images from the Wide Field Camera 3 (WFC3), which has not yet been used for weak-lensing studies. Both clusters appear to possess relaxed morphology in projected mass distribution, and their mass centroids agree nicely with those defined by both the galaxy luminosity and X-ray emission.  Using a Navarro--Frenk--White profile, for which we assume that the mass is tightly correlated with the concentration parameter, we determine the masses of SPT-CL~J2040--4451 and IDCS~J1426+3508
to be $M_{200}=8.6_{-1.4}^{+1.7}\times10^{14}~M_{\sun}$ and $2.2_{-0.7}^{+1.1}\times10^{14}~M_{\sun}$, respectively. The weak-lensing mass of SPT-CL~J2040--4451 shows that the cluster is clearly a rare object. Adopting the central value, the expected abundance of such a massive cluster at $z\gtrsim1.48$ is only~$\mytilde0.07$ in the parent 2500 sq. deg. survey. However, it is yet premature to claim that the presence of this cluster creates a serious tension with the current $\Lambda$CDM paradigm unless that tension will remain in future studies after marginalizing over many sources of uncertainties such as the accuracy of the mass function and the mass-concentration relation at the high mass end.
The mass of \idcs~is in excellent agreement with our previous Advanced Camera for Surveys-based weak-lensing result while the much higher source density from our WFC3 imaging data makes the current statistical uncertainty  $\mytilde40$\% smaller.
\end{abstract}

\altaffiltext{1}{Department of Astronomy, Yonsei University, 50 Yonsei-ro, Seoul 03722, Korea}
\altaffiltext{2}{Department of Physics, University of California, Davis, One Shields Avenue, Davis, CA 95616, USA}
\altaffiltext{3}{Korea Astronomy and Space Science Institute, 776, Daedeokdae-ro, Yuseong-gu, Daejeon 34055, Korea}
\altaffiltext{4}{University of Science and Technology, Daejeon 34113, Korea}
\altaffiltext{5}{Physics Division, Lawrence Berkeley National Laboratory, 1 Cyclotron Road, Berkeley, CA, 94720, USA}
\altaffiltext{6}{Department of Astronomy, University of Florida, Bryant
Space Science Center, Gainesville, FL 32611, USA}
\altaffiltext{7}{Department of Physics and Astronomy, University of Missouri,
5110 Rockhill Road, Kansas City, MO 64110, USA}
\altaffiltext{8}{Jet Propulsion Laboratory, California Institute of Technology, Pasadena, CA 91109, USA}


\section{INTRODUCTION} \label{section_introduction}

Galaxy cluster abundances have been well-accepted as powerful probes of cosmology because both the growth rate of clusters and the physical volume containing them are sensitive to cosmological parameters (e.g., Albrecht et al. 2006). Since longer-redshift baselines provide greater cosmological leverages, there have been relentless efforts to find and study clusters at higher and higher redshift  (e.g., Stanford et al. 2012; Stalder et al 2013; Muzzin et al. 2013; Tozzi et al. 2015; Wang et al. 2016).

Unfortunately, finding new galaxy clusters at high redshift is difficult. With the surface brightness declining as $(1+z)^{-4}$ and with average cluster masses
decreasing steeply with redshift, detecting clusters based on X-ray emission  becomes inefficient with redshift.
Complementary to X-ray surveys are deep infrared (IR) surveys, which rely on overdensity of galaxies against the background. In general, the samples obtained in this way are considered less biased toward massive clusters. Sunyaev-Zeldovich (SZ) observations have been hailed as very efficient tools to detect high-$z$ clusters because the $(1+z)^{-4}$ surface brightness dimming is cancelled by the $(1+z)^{4}$ boosting of the cosmic microwave background photon density at the cluster redshift.

In order to place meaningful constraints on cosmology, one needs to obtain reliable masses of clusters. Without sufficient understanding of gas physics in galaxy clusters, the conversion of observables in X-ray or SZ data will produce biased results. Similar issues are present when one uses cluster galaxy properties such as velocity dispersions or richness. Furthermore, high-redshift clusters are dynamically young, only recently detached from the Hubble expansion. Hence, application of hydrostatic equilibrium conditions to the sample is supposed to produce greater scatter/bias than when one studies low-$z$ clusters.

Weak lensing (WL) enables us to measure cluster masses without any dynamical assumptions. This  advantage becomes more critical when one probes dynamically young, high-redshift clusters. However, it is important to realize that weak-lensing studies face some difficulties for clusters at $z\gtrsim1$. First, signals get weakened as the redshift of a lens approaches that of source galaxies. Second, one has to extract subtle distortions from very faint and small galaxies, whose shapes are easily influenced by instrumental systematics such as anisotropic point-spread-functions (PSFs). Third, it is difficult to cleanly separate source galaxies based on limited colors because a significant fraction of members of clusters at high redshift in general are blue galaxies. Fourth, the signal becomes much noisier than in the case of low-$z$ clusters because the number of usable background galaxies decreases and the clusters in general become much less massive (thus creating fewer image distortions).

Space-based imaging provides great advantages when one measures weak-lensing signals for clusters at $z\gtrsim1$. Diffraction-limited PSFs allow us to measure subtle distortions of faint and small galaxy images with greater accuracy than turbulence-limited PSFs from the ground. In fact, to date most high-redshift galaxy clusters at $z\gtrsim1$ that have been analyzed with weak-lensing are based on {\it Hubble Space Telescope (HST)} imaging data. Currently, the most distant cluster that has been measured with weak-lensing is
\idcs~(Mo et al. 2016) at $z=1.75$, followed by XMMXCS~J2215-1738 at $z = 1.46$ (Jee et al. 2011).

The ``See Change" project is a large ($\mytilde174$ orbit) {\it HST} multi-epoch (Cycles 22 and 23) program to study 12 massive clusters at $z>1$ (PI. S. Perlmutter). The primary scientific goal is the enhancement of our knowledge on the expansion history of the universe via Type Ia supernova observations (e.g., Suzuki et al. 2012; Rubin et al. 2017) and on a mass function of massive clusters at the high end via weak lensing. We use the Wide Field Camera 3 (WFC3) to observe each cluster repeatedly with the filters F814W, F105W, F140W, and F160W. The optical WFC3-F814W imaging data are too shallow to resolve faint distant galaxies needed for a high-redshift cluster weak-lensing study.
On the other hand, the depth in the IR filters is high (exceeding $\mytilde28$th at the 5 $\sigma$ limit) in most cases. Therefore, we choose to measure shape distortions from WFC3-IR channel images, which have never before been used for weak-lensing studies.

In this paper, we report our first weak-lensing studies of two massive clusters, \idcs~and~\spt,~at $z\gtrsim1.5$ from the ``See Change" program. \spt~was discovered in the initial 720 sq. degree South Pole Telescope SZ (SPT-SZ) survey (Reichardt et al. 2013). The mass of the cluster inferred from the SZ data is $M_{200}=(5.8\pm1.4)\times10^{14} M_{\sun}$ (Bayliss et al. 2014). Given its redshift $z=1.48$, this mass is exceptionally high; adopting the central value, the expected abundance of such a massive cluster is slightly less than unity in the parent 720 sq. deg. survey. To date, no weak-lensing study has been carried out to confirm its mass.
Our weak-lensing study of \spt~provides an independent mass and thus allows deeper understanding of this interesting system. When completed, our study of all samples in the ``See~Change" program will establish an SZ--WL scaling relation at $z>1$.

\idcs~was discovered in the IRAC Distant Cluster Survey (IDCS) (Stanford et al. 2012). Brodwin et al. (2016) estimated the mass of the system to be $M_{200} = (4.1\pm 1.1)\times 10^{14} M_{\sun}$ based on the Combined Array for
Research in Millimeter-wave Astronomy (CARMA) SZ data.
Our first weak-lensing study of this cluster via Advanced Camera for Surveys (ACS) imaging gives a somewhat lower mass $M_{200}=2.3_{-1.4}^{+2.1}\times 10^{14} M_{\sun}$ (Mo et al. 2016). Although their  error bars overlap, the abundances estimated from the two mass measurements are different by a factor of $\mytilde24$.
Our weak-lensing study from this new independent (and higher signal-to-noise ratio (S/N) as we demonstrate in this paper) data set and pipeline will provide an important cross-check.

Throughout the paper, we assume the cosmology published in Planck Collaboration et al. (2016).
With the adopted set of the cosmological parameters, the plate scales are $\mytilde8.70$ kpc$~\mbox{arcsec}^{-1}$ and  $\mytilde8.71$ kpc$~\mbox{arcsec}^{-1}$
for \idcs~($z=1.75$) and \spt~($z=1.48$), respectively.
The $M_{200}$ value that we adopt here as a halo mass is a spherical mass within $r_{200}$, inside which the mean density becomes 200 times the critical density of the universe at the redshift of the cluster.

This paper is organized as follows. We describe our data, reduction, and weak-lensing procedure in \textsection\ref{section_observation}. The main results including cluster mass distribution and estimation are presented in \textsection\ref{section_result}. In \textsection\ref{section_discussion} we discuss non-statistical uncertainties in our cluster mass estimation from various sources and compare our weak-lensing masses with previous studies before we summarize in \textsection\ref{section_summary}.

\section{OBSERVATIONS}
\label{section_observation}
\begin{deluxetable}{ccc}
\tabletypesize{\scriptsize}
\tablecaption{HST data of \spt~and \idcs ~used in the current study.}
\tablenum{1}
\tablehead{\colhead{Galaxy cluster} &  \colhead{WFC3 filter} & \colhead{Exposure time} }
\tablewidth{0pt}
\startdata
\spt$^1$ & F814W & 5388 s \\
     & F105W & 16061 s \\
     & F140W & 17063 s \\
     & F160W & 6039 s \\
\hline
\rule{0pt}{2.5ex} \idcs$^2$ & F814W & 2468 s \\
      & F105W & 14863 s \\
      & F140W & 15916 s \\
      & F160W & 9057 s 
\enddata
\tablecomments{1. HST programs 13677 and 14327. 2. HST programs 11663, 12994, 13677, and 14327. }
\end{deluxetable}

\subsection{Data Reduction}
\spt~and \idcs~ were observed as part of the See~Change project (PROGRAM ID 13677 and 14327). The two clusters were imaged with WFC3-IR F105W, F140W, F160W, and WFC3-UVIS F814W filters. The choices for filters and cadence were designed to maximize the detection rate of well-measured high-redshift supernovae while minimizing the number of HST orbits. The cluster \idcs~was also observed under the programs 11663 (PI. M. Brodwin) and 12994 (PI. A. Gonzalez), and we utilize the images from these programs to improve the depth. We summarize the resulting exposure times in Table~1.

Our reduction starts with the output (FLT or FLC images) generated by the STScI {\tt calwf3} pipeline, which removes instrumental signatures except for geometric distortion. The current version (v3.3) also automatically corrects for charge transfer efficiency (CTE) degradation problems in the UVIS detector via the PCTECORR step (Bajaj 2016). In our previous studies with the ACS, rigorous analysis on the fidelity of the correction mechanism had to be demonstrated because the CTE-induced charge trailing directly impacts galaxy shape measurement (e.g., Jee et al. 2014a). In the current study, we measure weak-lensing signals from WFC3-IR images, which do not suffer from CTE degradation. Hence, we do not present our fidelity verification of the CTE correction in this paper. The FLT/FLC images are cosmic ray-rejected and combined using the MultiDrizzle package after we estimate relative shifts between images from common astronomical sources. The final pixel scale is chosen to be 0.05$\arcsec~\mbox{pix}^{-1}$. This scale is selected in order to mitigate the undersampling artifact of the WFC3-IR PSF; the FWHM of WFC3-IR is $\mytilde0.141\arcsec$ at $\lambda=14000$\AA, which is similar to the
native pixel scale $\mytilde0.13\arcsec$ of the WFC3-IR detector.
The final pixel scale (0.05$\arcsec~\mbox{pix}^{-1}$) is slightly larger than the native pixel scale 0.04$\arcsec~\mbox{pix}^{-1}$ of WFC3-UVIS. However, this is not a concern for our science because we 
do not use the WFC3-UVIS F814W image for our weak-lensing study.
The drizzling kernel is chosen to be ``gauss" with the {\tt pixfrac} parameter set to 0.7.   

We combine F105W, F140W, and F160W filters to create one deep detection image. Sources are detected with SExtractor in dual image mode by looking for at least five connected pixels whose values are above 1.5 times the sky rms. Because of the depth, the resulting object densities are very high ($\mytilde670~\mbox{arcmin}^{-2}$ and $\mytilde630~\mbox{arcmin}^{-2}$ for \spt~and~\idcs, respectively).

\subsection{Notes on Potential Weak-lensing Systematics Due to WFC3-IR Detector Features}

The WFC3-IR detector uses a 1024$\times$1024 HgCdTe photovoltaic array, which is photosensitive within the 4000\AA$\sim$17000\AA~ wavelength range (Dressel et al. 2016). The effective  lower limit in wavelength is set by both the IR channel filters ($\gtrsim9000$\AA) and the detector coating ($\gtrsim10000$\AA). Of the most striking differences of this IR detector from conventional CCDs is its ability to read each pixel non-destructively multiple times. Certainly, this feature provides  a clear advantage over CCDs because it does not suffer from CTE degradation, which systematically
elongates astronomical objects along the readout direction and thus makes HST weak-lensing studies based on CCD images non-trivially cumbersome (e.g., Jee et al. 2014a). However, the new instrument also possesses different systematics that may have potential implications for weak-lensing. Here, we discuss what these potential systematics are and how much they matter for the current weak-lensing study.

\subsubsection{Interpixel Capacitance (IPC)} 
The charge collected by an individual pixel is not entirely localized within the pixel in many IR detectors, and the WFC3 IR detector is no exception. The resulting effect appears as the  diffusion of the charge to the pixel's nearest pixels. This effect is similar to charge diffusion in CCDs, where clouds of charges spread before being collected in potential wells. However, in IR detectors such as the WFC3-IR channel, the spreading happens because of a deterministic cross-talk (capacity coupling) between pixels (Brown et al. 2006). Thus, unlike charge diffusion, the charge
spreading due to interpixel capacitance is not color-dependent.
The on-orbit measurement by
Hilbert \& McCullough (2011) shows that about 6.3\% of the flux in a single pixel source is symmetrically re-distributed to the neighboring eight pixels. They also demonstrate that the result can be described by a deterministic convolution with a 3$\times$3 kernel without any clear dependence on flux, color, or time. Although this simple mathematical model may suggest a possibility to use deconvolution to remove IPC (McCullough 2008), the current STScI pipeline does not automatically correct for the effect. In this study, we choose to remove the IPC effect by letting our PSF model include the broadening by IPC. This is similar to the approach in our previous weak-lensing studies based on ACS data, where we did not separately model the charge diffusion, but instead used the stellar PSF that already included the charge diffusion. Because the IPC effect is more deterministic than charge diffusion, 
the current approach with the WFC3-IR imaging data should create fewer systematic errors.

\subsubsection{Persistence} 
While IR detectors do not suffer from charge bleeding, pixels subject to high signal levels show higher dark-current levels in subsequent exposures.
This phenomenon is called ``persistence" and manifests itself as afterglows from earlier exposures. 
The physical mechanism leading to persistence is common to many IR detectors based on HgCdTe and is well understood. Long et al. (2012) find that the persistence in the WFC3-IR detector is a nonlinear function of the fluence (total number of electrons generated in a previous exposure) and the history that the pixel is held at high fluence levels. There are some guidelines on strategies to minimize persistence and also even some phenomenological models to remove the effects (Dressel et al. 2016). However, since this quantum mechanical phenomenon is very sophisticated, Dressel et al. (2016) warn that the model is still immature. A large amount of the WFC3-IR persistence becomes noticeable only for pixels whose fluence levels are close to the saturation, and thus the number of pixels significantly affected by persistence is expected to be small in most cases. In our study, in order to examine whether or not our IR imaging data are significantly affected by persistence, we utilize the STScI tool\footnote{https://archive.stsci.edu/prepds/persist/search.php} which estimates the persistence for our data sets based on previous observations. We verify that the predicted persistence level is usually low;
less than about $0.1$\% of the detector area has the dark current rate $\gtrsim0.01~\mbox{e}^{-}s^{-1}$ for both clusters. However, for several exposures of \idcs~(e.g., the dataset ic3b04s1q), we find that the fraction of the pixels with the same dark current rate increases by a factor of 10 or higher. Our visual inspection of these exposures reveals that a number of large-scale low surface brightness streaks are present. These streaks are found to be the afterglows of the previous calibration observation of dark Earth's airglow (PROG ID 13068). In the current study, we remove all WFC3-IR pixels whose predicted persistence level is higher than $0.01~\mbox{e}^{-}\mbox{s}^{-1}$ by flagging them as bad pixels. 

\subsubsection{Detector Nonlinearity} 
In addition to the IPC and persistence issues discussed above, another importance feature of the HgCdTe detector is nonlinearity. Dressel et al. (2016) report that without any correction the level of nonlinearity is at the 5\% level when the count reaches $\mytilde94$\% of the full saturation. A potential issue of the nonlinearity for weak-lensing analysis is its distortion of PSFs. Because we model PSFs using bright stars, there are chances that the model PSF has a less sharp core than the actual PSF relevant to faint galaxies which contain weak-lensing signals. The expected impact of this PSF discrepancy on galaxy shear measurement might be multiplicative, which perhaps leads to slight overestimation of shear because the model PSF would be wider than the galaxy PSF. A potentially useful method to estimate the shear calibration error arising from this effect is investigation of PSF width as a function of flux utilizing stellar field imaging data. In the current study, however, we choose to address the issue simply by avoiding selection of too bright stars in our PSF model construction. As mentioned above, the reported nonlinearity is at the 5\% level when the count reaches $\mytilde94$\% of the full saturation. The multiplicative shear calibration error caused by this level of nonlineary is expected to be $\mytilde1$\%, according to our simplified image simulation where we model both galaxies and PSFs with a Gaussian profile.
This level of bias is not relevant in our cluster lensing, where the dominant source of error is statistical ($>10$\%).

\subsubsection{Undersampling} 
\label{section_undersampling}
The WFC3-IR detector undersamples the PSF by a factor of two; the FWHM of the F140W PSF is $\mytilde1.1$ pixel. Although one can attempt to mitigate the resulting effect by dithering and combining many images with a new smaller target pixel scale, the information destruction from the binning of the native pixel scale is to a large extent irreversible. This causes several complications in weak-lensing analysis including PSF modeling and galaxy shape measurement. The PSF core is flattened and the resulting shape distortion is sensitive to where the source center lies within a pixel. This PSF distortion also affects extended sources unless their sizes are sufficiently larger than that of the PSF. In fact, in weak-lensing studies of high-$z$ clusters, the signal mostly comes from faint, small galaxies which are only marginally larger.

Perhaps the most promising method to resolve the issue is to forward-model the observed images of PSFs and galaxies while simulating the procedure leading to the distortion caused by undersampling. This requires modeling both PSF and galaxy profiles at high-resolution (much higher than the native detector pixel scale) and applying two-dimensional binning to generate/simulate observed WFC3 images. These ``model" observed images can be repeatedly compared with the ``real" observed images until the residuals become sufficiently small. One can expand this forward-model approach so that the model may include other features of the instrument such as IPC.

However, in this study, we use our existing weak-lensing pipeline and decide to calibrate the undersampling bias (and other sources of biases) by utilizing external ACS data. The decision 
was made primarily because the forward-model approach discussed above is expensive. In addition, 
even if the forward-model pipeline is implemented, the shear calibration through simulation or external data is still inevitable because there exist other sources of bias (e.g., noise bias, model bias, etc.) that the forward-model approach cannot adequately address. Our shear calibration procedure through external ACS data is described in \textsection\ref{section_shear_calibration}.

\subsection{PSF Modeling}

PSFs both dilute and bias ellipticity measurements at the same time. Dilution happens because the magnitude of galaxy ellipticities on average decreases by the smearing effect of the PSF while the orientation of galaxy ellipticities tends to be aligned with the PSF anisotropy.
Although HST PSFs are much smaller than those from ground-based telescopes, their impact on galaxy shapes are still non-negligible when one carries out quantitative weak-lensing studies. As in our previous studies, we utilize archival images of star clusters to sample and construct PSF models. The images that we consider are drawn from the program 11453, which targets relatively sparse regions of the globular cluster 47~Tuc. The observation was originally planned to derive low-frequency flat fields of the IR detector.

\begin{figure}
\centering
\includegraphics[width=9cm,trim=0cm 4cm 0cm 0cm]{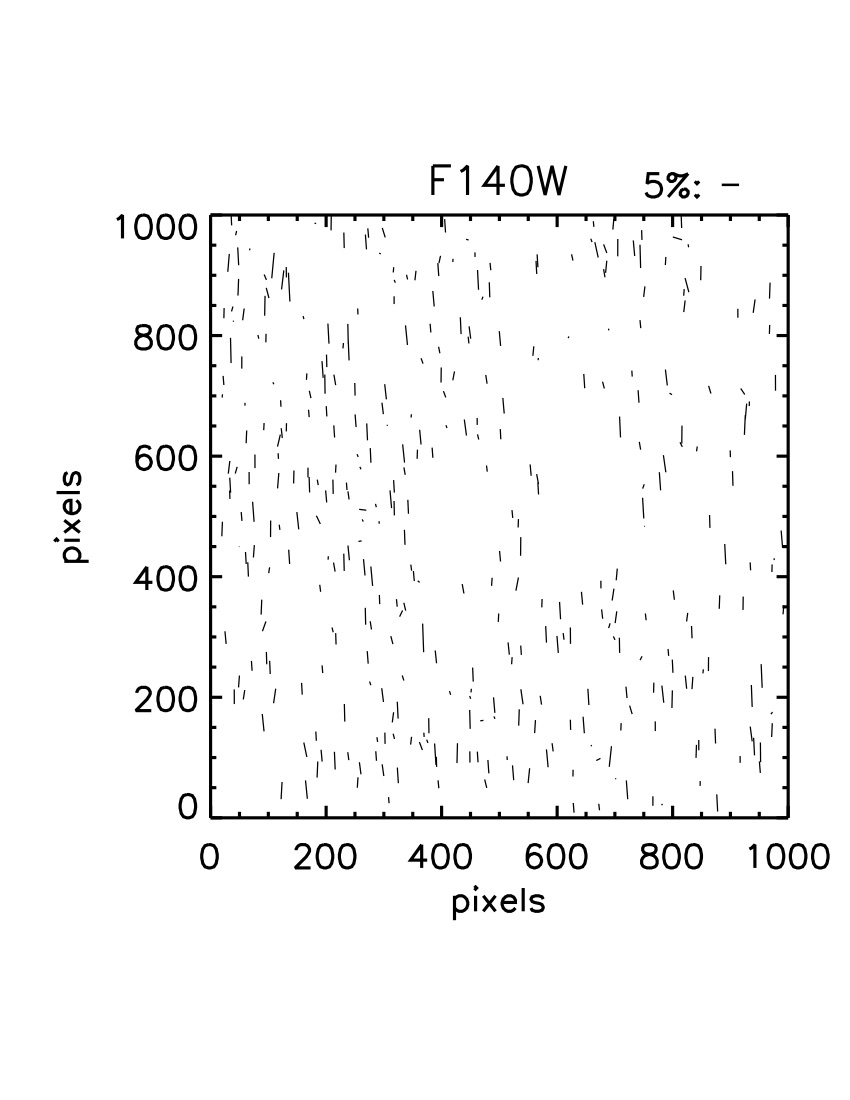}
\caption{Uncorrected WFC3-IR PSF elongation due to geometric distortion caused by the focal plane tilt. Shown here is drawn from the F140W image of 47~Tuc. The length and direction of the sticks represent the magnitude and orientation of ellipticity, respectively (the horizontal stick above the plot window indicates the length of 5\% ellipticity). In the image-based orientation that we use here, the PSFs are elongated vertically. The average ellipticity is $\mytilde3\%$, which is consistent with our expectation; the $\mytilde7$\% geometric distortion translates to $\mytilde3$\% ellipticity according to the adopted ellipticity definition in this study.}
\label{fig:F140W_psf}
\end{figure}

The WFC3 is located at the principal telescope axis, and thus one might expect a relatively small geometric distortion relative to the ACS, which is $\mytilde6\arcmin$ off-axis on the focal plane. However, because of the tilted focal plane, in fact, the geometric distortion is significant, projecting a square into a rhombus for the UVIS detector and into a rectangle for the IR detector. Accordingly, in the IR channel, from which here we desire to measure galaxy shapes for weak-lensing, the PSFs are elongated vertically in the image-based coordinate system.
Figure~\ref{fig:F140W_psf} shows such an example drawn from the data sets taken on 2009 July 13.
We observe similar patterns for different filters of WFC3-IR regardless of observation epoch. 

\begin{figure}
\includegraphics[width=9cm,trim=0cm 4cm 0cm 0cm]{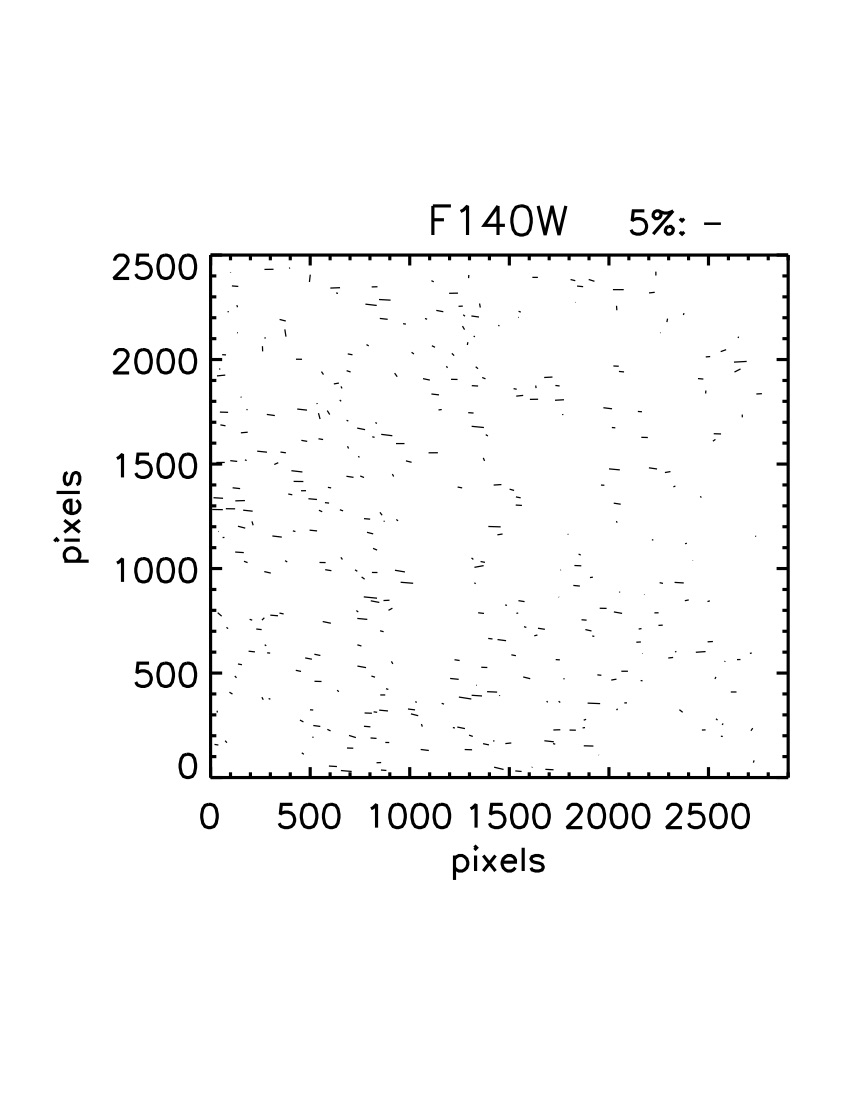}
\caption{Same as in Figure~\ref{fig:F140W_psf} except that this time the analysis is carried out with the geometric-distortion corrected image. We use a final pixel scale of 0.05$\arcsec$/pixel with {\tt pixfrac=0.7} to ``drizzle" the image. After distortion correction, the resulting image becomes elongated in the x-axis as shown. It is clear that not only the vertical elongation disappears, but also there exists residual PSF anisotropy in the horizontal direction at the $\mytilde1$\% level. It is very unlikely that this pattern is caused by overcorrection of the geometric distortion using the model of Kozhurina-Platais et al. (2012) because the amount of the overcorrection has to be as large as  $\mytilde2$\% in order to produce the observed level of residual PSF ellipticities.}
\label{fig:F140W_psf_corr}
\end{figure}

Kozhurina-Platais et al. (2012) characterize geometric distortions of the WFC3 with fourth order polynomials by comparing $\omega$~Cen field data of the WFC3 with
the ACS standard catalog. They estimate that their correction is accurate within $\mytilde7$ mas ($\mytilde0.1$ pixels). Because the solution of  Kozhurina-Platais et al. (2012) is included in the MultiDrizzle package, we can readily test the fidelity of the geometric distortion model by processing (correcting geometric distortion of) the 47~Tuc images with Multidrizzle and repeating the above star-ellipticity analysis. 
Since we create our scientific images of the two galaxy clusters using a final pixel scale of 0.05$\arcsec$/pixel with {\tt pixfrac=0.7}, we match the ``drizzling" parameters of the 47~Tuc images to these values. One of our test results is displayed in Figure~\ref{fig:F140W_psf_corr}.  
It is easy to see that the vertical elongation pattern in Figure~\ref{fig:F140W_psf} disappears when the distortion-corrected image is used. However, a close examination reveals that the residual PSF elongation is in fact horizontal in many cases. These residual ellipticities should be related to optical aberration, rather than overcorrection of the focal plane tilt unless the geometric distortion model of Kozhurina-Platais et al. (2012) overcorrects the distortion by as much as $\mytilde2$\%, which is very unlikely.

Following our previous HST-based weak-lensing studies with the ACS, we decompose the WFC3-IR PSFs on distortion-corrected images using principal component analysis (Jee et al. 2007a). Similarly to the ACS PSFs, we find that the first 20 principal components are responsible for $\mytilde95$\% of the total variance.
We employ third-order polynomials to interpolate the PSFs and model the variation across the WFC3-IR detector. Readers are referred to our previous ACS-based studies for details on how these PSF models are used to describe the PSF pattern in weak-lensing fields (e.g., Jee et al. 2011).

\subsection{Shear Measurement} \label{section_shape_measurement}

Gravitational lensing induces a small change in object ellipticity relative to the intrinsic shape dispersion. Therefore, we must average over ellipticities from a sufficient number of sources to measure gravitational shear. How one defines ellipticity for each source is a subjective matter because galaxy morphologies are complex and diverse (often possessing radially varying ellipticities).
Our method determines ellipticity by fitting an elliptical Gaussian to a galaxy that minimizes the residual when the model is subtracted from the galaxy image. Of course, we must convolve the elliptical Gaussian with the model PSF expected at the galaxy position to take into account PSF effects which, when uncorrected, give smaller ellipticities and also bias galaxy orientation toward the direction of the PSF elongation. The semimajor and semiminor axes, $a$ and $b$, are used to define our ellipticity $e=(a-b)/(a+b)$. Practically, one should use two components to represent ellipticity with direction. Therefore, we define $e_1=e\cos 2 \phi$ and $e_2=e\sin 2 \phi$, where $\phi$ is the position angle of the major axis of the ellipse measured counter-clockwise from the positive $x$-axis.

Reduced shears $g=\gamma/(1-\kappa)$ are estimated by weight-averaging galaxy ellipticities and applying calibration factors as follows:
\begin{equation}
g_{1(2)} = m_{1(2)} \frac{1}{W} \sum_{i=1}^{N} e_{1(2)}^{\prime} \mu_i
\end{equation}
\noindent
where $\mu_i$ is the inverse-variance weight:
\begin{equation}
\mu_i  = \frac {1} { \sigma_{SN}^2 + (\delta e_i)^2}, \label{eqn_shear_weight}
\end{equation}
\noindent
$W$ is $W=\Sigma \mu_i$, and $m_{1(2)}$ is the multiplicative bias factor, which is required to reconcile
the difference between measured ellipticity and theoretical (reduced) shear.

\subsection{Source Galaxy Selection}
\label{section_source_selection}

Gravitational lensing occurs when light bundles from background galaxies are deflected by foreground masses, and hence care must be taken when selecting a source population. An ideal situation might be the case when redshifts are known accurately for all sources and we simply select the galaxies whose redshifts are greater than the lens. In reality, it is impossible to obtain accurate (i.e., spectroscopic) redshifts for all detected sources. A photometric reshift is in general the most useful alternative to a spectroscopic
redshift. However, since the requirements for depth and breadth in filter coverage are very high, the observation is often prohibitively expensive. Thus, in most cases of weak gravitational lensing studies, the source population is selected based on the information available from a small number of filters. 

In defining a source selection function based on broadband photometry,
two competing factors are purity and shot noise. In general, the lensing S/N value increases when purity 
becomes high (i.e., the fraction of contaminants becomes low). However, one's blind attempt to increase purity too aggressively often results in significant reduction of the total number of source galaxies, which decreases the overall lensing S/N  because of increased shot noise.
Thus, ignoring lensing efficiency of individual source redshifts, the S/N increases linearly with purity and the square root of source density:
\begin{equation}
S/N_{lens}\propto f_p \sqrt{n_s},
\end{equation}
\noindent
where $f_p$ is the purity ($1--$~contamination fraction) and $n_s$ is the source density.

Obviously, this statement is not entirely true when one considers the merit of faint sources, which are on average more distant and subject to greater lensing distortion.
The issue becomes even more complicated when one considers the ``quality" of shape. Lensing signals from faint galaxies near the limiting magnitude of the survey or small objects approaching the size of the PSF are somewhat diluted on average compared to those from well-resolved, bright galaxies. This effect is often broadly termed ``noise bias."

\begin{figure*}
\centering
\includegraphics[width=8.8cm]{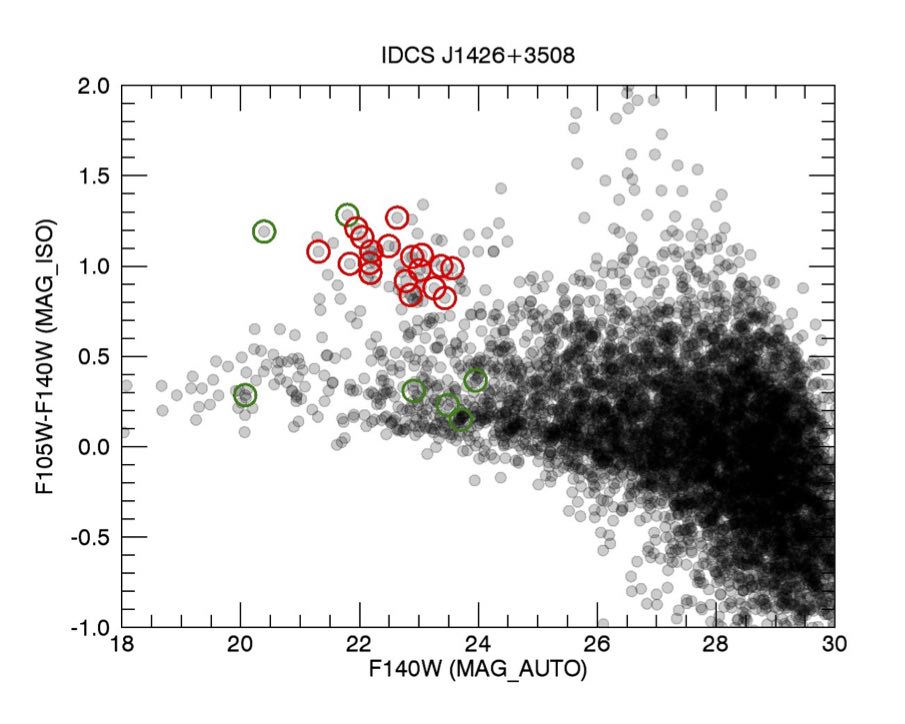}
\includegraphics[width=8.8cm]{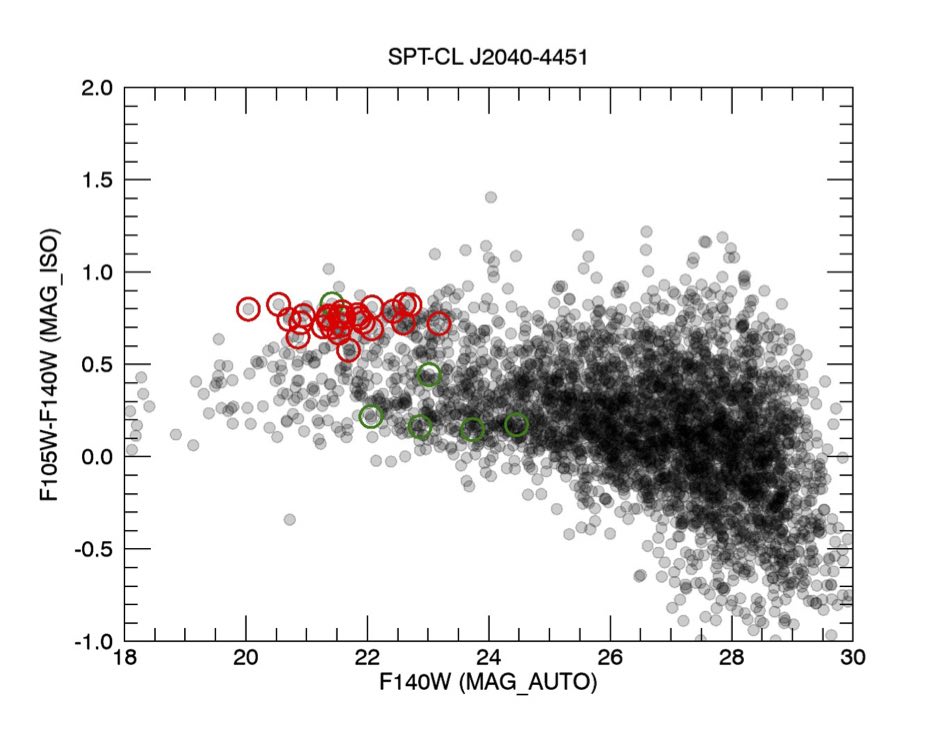}

\caption{Color-magnitude relation in \idcs~and \spt. The 4000\AA~ break feature redshifted to $\mytilde11000$\AA~ and $\mytilde9920$\AA~ at the cluster redshifts $z=1.75$ and $1.48$, respectively, are bracketed by the  F105W-F140W color. Hence, ``passive" galaxies in both clusters should occupy a distinct locus in this color-magnitude diagram.
Green circles represent the colors of the spectroscopic members. For \idcs, all seven members reported in Stanford et al. (2012) are found within the area covered by our ``SeeChange" HST program. For \spt, on the other hand, only six out of the 15 spectroscopic members in Bayliss et al. (2014) are covered. 
Because both Stanford et al. (2012) and  Bayliss et al. (2014) used the emission lines (sensitive to star formation rate) to determine the membership, these galaxies are not passive in general. Red circles represent the member candidates that we identify based on their early-type morphology and F105W-F140W color at F140W$\lesssim23.5$. When we
perform synthetic photometry by redshifting the Elliptical galaxy template of Coleman et al. (1980) to $z=1.75$ and $1.48$, the expected F105W-F140W colors at these redshifts are $\mytilde1.38$ and $\mytilde0.97$, which are slightly ($0.1-0.2$) redder than the observed colors of the candidates.
}
\label{fig_cmd}
\end{figure*}

In this study, we consider the factors discussed above empirically similarly to our previous methods. Point sources are removed based on their half-light ratios. Also, extremely small galaxies are excluded by requiring that the semi-minor axes of surviving sources be greater than 0.3 pixels after our PSF-effect removal.
Cluster red-sequence
galaxies are identified based on their F105W-F140W colors; the 4000\AA~ break feature is redshifted to $\mytilde11000$\AA~ and $\mytilde9920$\AA~ at the cluster redshifts $z=1.75$ and $1.48$, respectively. In Figure~\ref{fig_cmd}, we show the color magnitude diagrams of the two clusters. It is easy to verify the enhancement of galaxy density in the particular region of the F105W-F140W versus F140W space. Blind to these color magnitude diagrams, we also identify some bright early-type galaxies from visual inspection based on their morphology and color using our color-composite images (created from F814W, F105W, and F140W). We verify that these independently selected galaxies closely trace the expected red-sequence locus in these color magnitude diagrams. For our source selection, we discard objects whose F104W-F140W colors are greater than 0.5 and 0.7 for \spt~and~\idcs, respectively. The amplitude of lensing signals varies as we adjust these color-cut values, although not to the extent that an optimization is needed. 
The bright magnitude limit is set to F140W=24 for both clusters. If we lower the limit below the cut F140W=24 (i.e., including more bright galaxies), the overall lensing S/N value decreases because the purity loss due to contamination starts to offset the reduction in shot noise. 

The faint limit is imposed indirectly by requiring that ellipticity measurement error be less than $0.25$, which results in F140W$\lesssim28$  for both clusters. Including fainter galaxies decreases the overall S/N while an opposite trend is observed when we exclude more galaxies by tightening the ellipticity error criterion.
The final source number density is $\mytilde240~\mbox{arcmin}^{-2}$ for both clusters. This  value is approximately a factor of two higher than the typical numbers that we achieved in our previous ACS weak-lensing studies (Jee et al. 2011).

Quantitative interpretation of the lensing signals becomes possible if we know the redshift distribution of the source galaxies along with shear calibration. Our shear calibration is described in \textsection\ref{section_shear_calibration}, and here we discuss the procedure for redshift estimation. For ACS weak-lensing studies, we have used the Coe et al. (2006) photometric redshift catalog. The catalog was generated 
for the Ultra Deep Field (UDF; Beckwith et al. 2006)
prior to the installation of the WFC3 and thus lacks its photometric data. Although Coe et al. (2006) computed photometric redshifts utilizing very deep NICMOS imaging data available at the time, for the current analysis based on WFC3-IR imaging data, the catalog is less than optimal. Hence, in this paper, 
we use the new photometric
redshift catalog reported in Rafelski et al. (2015). They produce a high-quality photometric redshift catalog for the UDF from eleven passband data covering UV to near IR. While the old NICMOS imaging covers approximately half of the ACS UDF, this new catalog is obtained from the combination of both the UDF09 (PROG ID 11563, PI: G. Illingworth) and UDF12 (PROG ID 12498; PI: R. Ellis) campaigns 
with CANDELS (PROG IDs 12060, 12061, and 12062; PI. S. Faber and H. Ferguson) data, which provide the IR coverage for the entire UDF.

In weak lensing, the redshift distribution of the source population is often expressed
with the following $\beta$ parameter:
\begin{equation}
\beta  = \left < \mbox{max} \left (0,\frac{D_{ls}}{D_s} \right ) \right >,
\label{eqn_beta} 
\end{equation}
\noindent where $D_{ls}$ and $D_s$ are the angular diameter distances
between the lens and the source and between the observer and the
source, respectively.  
This $\beta$ parameter is used to compute the critical
surface density $\Sigma_c$ of the cluster given by

\begin{equation}
\Sigma_{\rm crit} = \frac{c^2}{4 \pi G D_l \beta }, \label{eqn_sigma_c}
\end{equation}
\noindent where $c$ is the speed of light, $G$ is the gravitational
constant, and $D_l$ is the angular diameter distance to the lens.
We apply the same color and magnitude selection criteria to the Rafelski et al. (2015) catalog; because the catalog does not include ellipticity measurement errors, we directly use F140W magnitudes to define the faint limit of the source population. Without any depth correction applied, the resulting $\beta$ values are 0.100 and 0.149 for \idcs~and~\spt, respectively. These values are biased high because the UDF is much deeper than our cluster fields. Therefore, it is important to count galaxies differently according to the magnitude discrepancy between our cluster fields and the UDF. The bias-corrected values are
 0.086 ($z_{eff}=2.05$\footnote{The effective redshift $z_{eff}$ is defined to be the source redshift yielding the $\beta$ value in Equation~\ref{eqn_beta}. This is different from the mean redshift of the sources because in the evaluation of $\left<\beta\right>$ we assign zero to the sources whose redshifts are lower than the cluster. }
 ) and 0.120 ($z_{eff}\simeq1.83$) for \idcs~and~\spt, respectively. 
We estimate that about 56\% of the source population is non-background in \idcs~whereas the contamination fraction is $\mytilde45$\% in \spt. Because the lensing kernel is nonlinear, the width of the distribution must be estimated to reduce the bias arising from the use of a single source plane $\left<\beta \right>$ (Seitz \& Schneider 1997). We obtain $\left<\beta^2 \right>$=0.034 and 0.052 for \idcs~and \spt, respectively.

\subsection{Cluster Member Contamination}
The redshift estimation of the source population in \textsection\ref{section_source_selection} assumes that the contamination to the source catalog by blue cluster members is negligible. If found significant, however, the contamination would dilute the lensing signal and lead to non-negligible underestimation of the cluster masses. Below we discuss this issue.

Many authors have argued that, when a limited number of filters are used in source selection by removing red-sequence galaxies, one of the important systematic errors is the contamination by faint blue cluster members (e.g., Broadhurst et al. 2005; Okabe et al. 2010; Applegate et al. 2014; Melchior et al. 2017). Recently, Medezinski et al. (2017) suggested that the blue cluster member contamination is not only limited to the cluster central region, but also to outskirts and will thus impact determinations of both mass and concentration. Since blue cluster members occupy a large volume in color-magnitude space (Ziparo et al. 2016), it is difficult to efficiently remove them by adjusting the color selection window. Furthermore, both observations and theories show that the fraction of blue cluster members increases with redshift, suggesting that the issue may become particularly critical in our analysis, which studies two high-redshift clusters at $z>1.5$.

When sources are selected photometrically based on a few filters, one can only attempt to estimate the contamination statistically. Typically, this is done  by comparing the number density of the sources with that obtained from some reference fields (e.g., Jee et al. 2005), by
examining the radial dependence of the source density (Medezinski et al. 2017), and/or by quantifying dilution of the lensing signal as a function of source selection (Broadhurst et al. 2005). In this paper, we use the first two methods. 

\begin{figure}
\centering
\includegraphics[width=8.8cm]{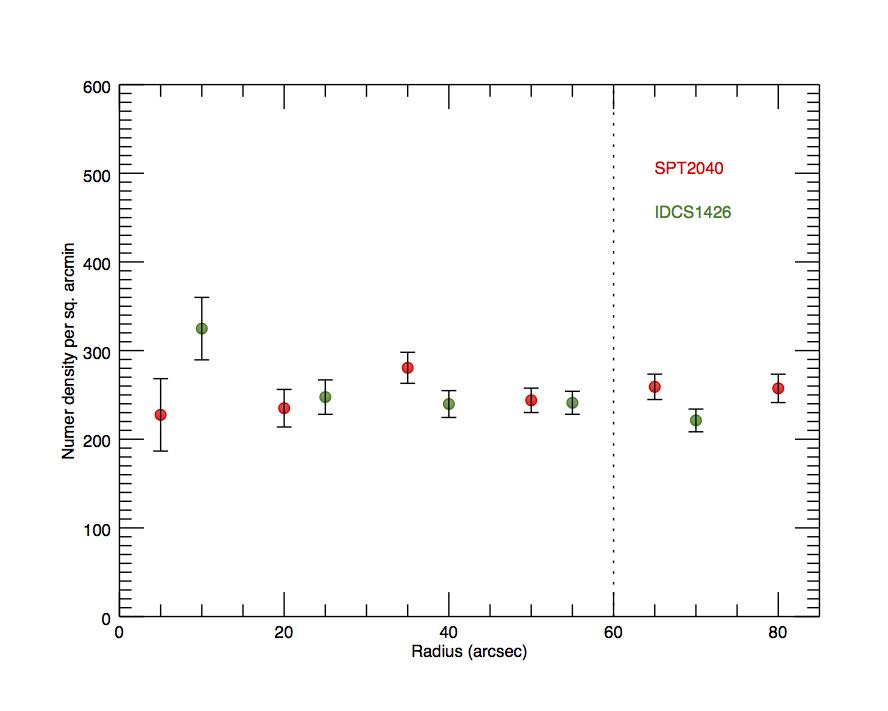}
\caption{Radial source density profile. We examine source densities as a function of radii from the cluster center (X-ray peak) for both clusters studied here. The annuli beyond $r\gtrsim60\arcsec$ (dotted line) cannot complete circles. For \spt~ there is no indication that the source density varies with radius. The source density profile of the \idcs~ field shows a slight excess at the smallest radius ($r\sim10\arcsec$). }
\label{fig_radial_density}
\end{figure}

Figure~\ref{fig_radial_density} shows the source density radial profiles in our cluster fields. This type of plot is often used to
detect cluster galaxy contamination under the assumption that the density enhancement near the cluster center should make the profile decline as a function of radius. However, the interpretation requires caution because there are many other factors affecting the shape of the profile such as  masking of background galaxies by the cluster members, magnification by the cluster potential, blending with the cluster galaxies, selection effects, etc.
With these caveats, we find no indication that the source density varies with radius for \spt. The source density profile of the \idcs~ field shows a slight excess at the smallest radius ($r\sim10\arcsec$). However, the significance is weak.

\begin{figure}
\centering
\includegraphics[width=8.8cm]{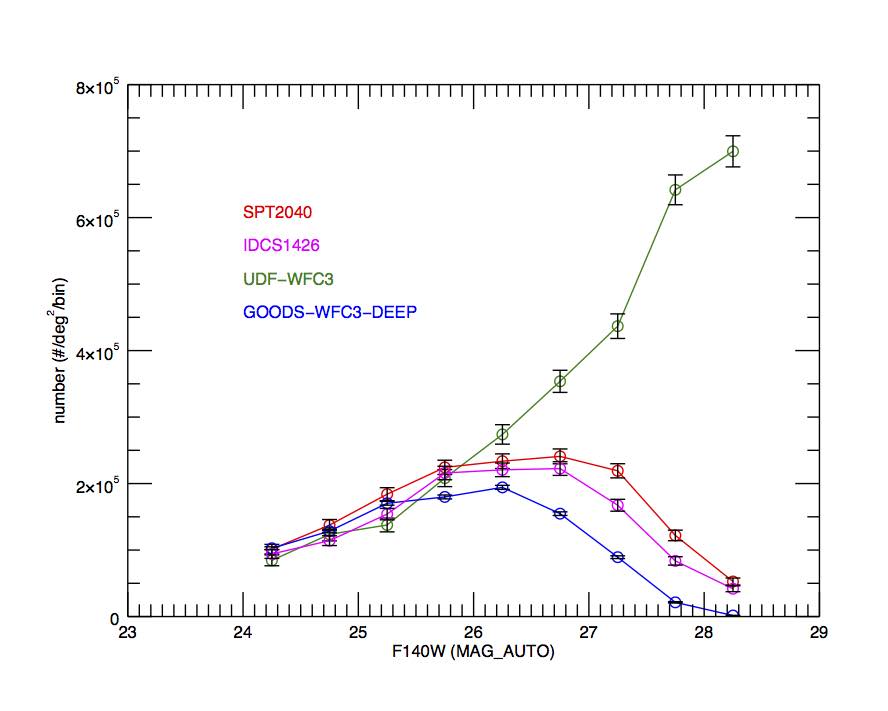}
\caption{Comparison of magnitude distributions between source and control populations. We use the deep WFC3 imaging data within the UDF and GOODS-S as control fields in order to examine source density contamination in the cluster fields. Only poissonian errors are included in the error bars. Since the GOODS-S does not have F140W, we perform photometric transformation using the F160W photometry. We apply the same source selection criteria to the galaxies in the control fields. At F140W$\gtrsim26$ the difference in galaxy density increases between the two control fields because of the depth discrepancy. We do not observe any significant source density excess at F140W$\lesssim26$. 
}
\label{fig_source_density}
\end{figure}

Figure~\ref{fig_source_density} displays the magnitude distributions of the source galaxies in the cluster fields compared with those from the two control fields. One control field is the GOODS south WFC3 deep field (Guo et al. 2013), which covers the middle third of the
GOODS-S ACS region with an area of $\mytilde$55 arcmin$^2$. Because of the relatively large area and depth, this field may serve as a fair control field for relatively bright sources when the sample variance is considered. However, because the depth is shallower (six orbits in F160W) than our cluster fields, the incompleteness is a limiting factor at F140W$\gtrsim26$. Therefore, we chose the HUDF12 WFC3 deep field (Ellis et al. 2013) as our second control field which, despite its small field area $\mytilde4.6~$arcmin$^2$, provides a better representation of the magnitude distribution at F140W$\gtrsim26$. We apply the same source selection criteria to both control field catalogs; since the GOODS data do not have F140W, we perform photometric transformation to convert the F160W mag to the F140W mag. In the  F140W$\lesssim26$ regime, the source magnitude distributions in the cluster fields are statistically consistent with those obtained from the control fields even if we ignore the sample variance, which might account for $\mytilde10$\% difference in the small cluster fields\footnote{We obtained this number by examining the source density of all SeeChange clusters. The results will be published in our future summary paper.}
In the  fainter (F140W$\gtrsim26$) regime, a direct comparison is impossible because both the cluster and the GOODS south data are incomplete. Nevertheless, when we degrade the HUDF12 data to match the depth of the cluster imaging data, the resulting magnitude distribution becomes also consistent with the cluster source distributions. Therefore, together with the radial source density profile experiment, we conclude that the blue cluster contamination (although it is surely present) is not a significant source of systematic errors in our lensing analysis. We reported similar findings in our previous high-$z$ cluster studies (e.g., Jee et al. 2005; Jee et al. 2011; Jee et al. 2014b).

\section{RESULTS}
\label{section_result}
\subsection{Two-dimensional Mass Distribution}
\label{section_mass_reconstruction}
A number of algorithms have been suggested in the literature to optimally convert a shear field into a two-dimensional mass density map. In this study, we use the MAXENT code of Jee et al. (2007b), which regularizes the result with the maximum entropy principle. In brief, the algorithm ensures that the solution converges to a case where we maximize the entropy of mass pixels. In other words, given the constraints set by galaxy shapes, we look for the smoothest possible solutions. Readers are referred to the original paper for algorithmic details.

\begin{figure*}
\includegraphics[width=9cm]{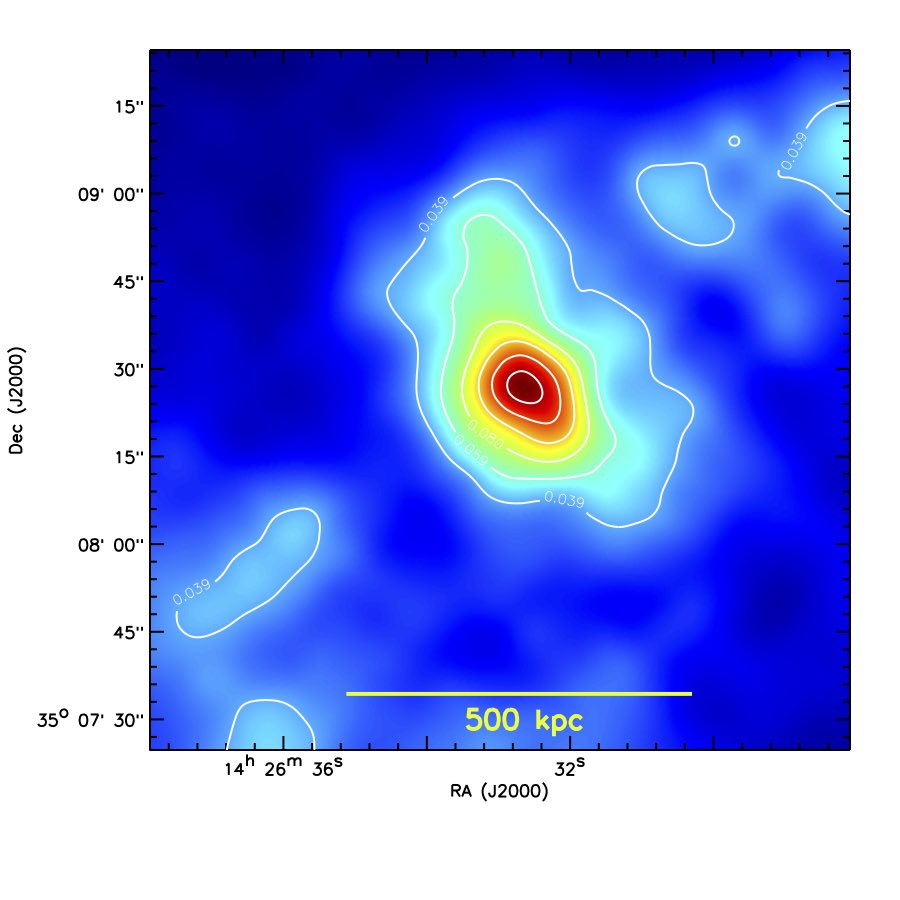}
\includegraphics[width=9cm]{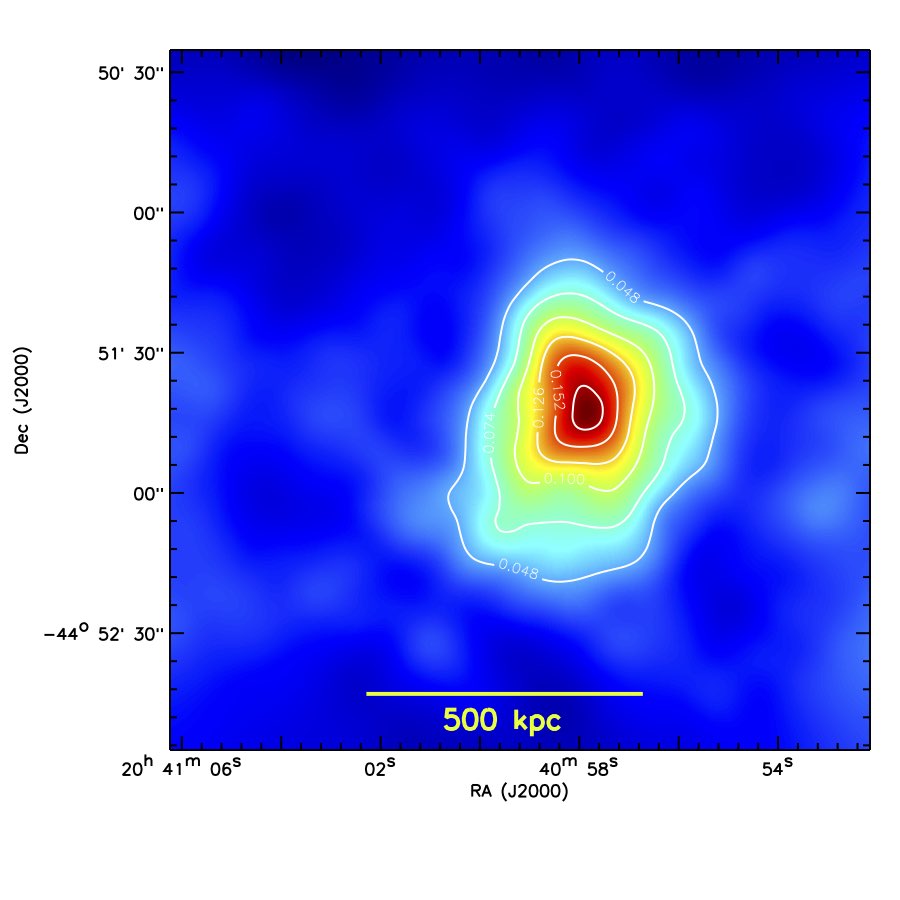}
\includegraphics[width=9cm]{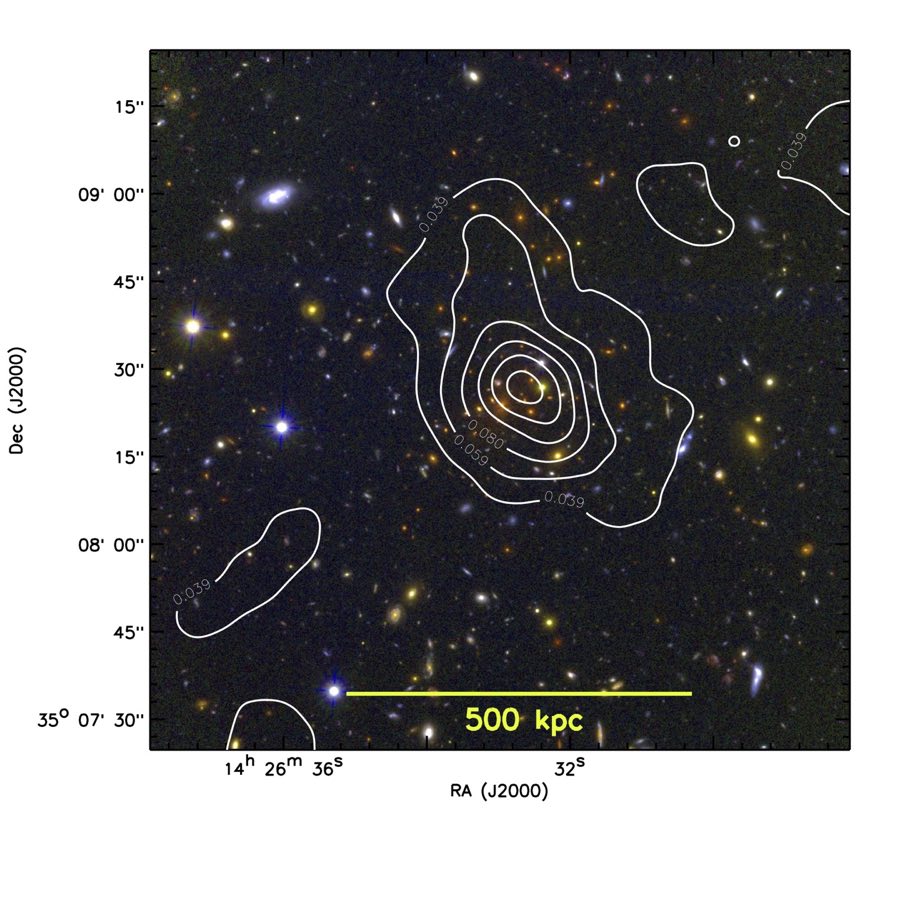}
\includegraphics[width=9cm]{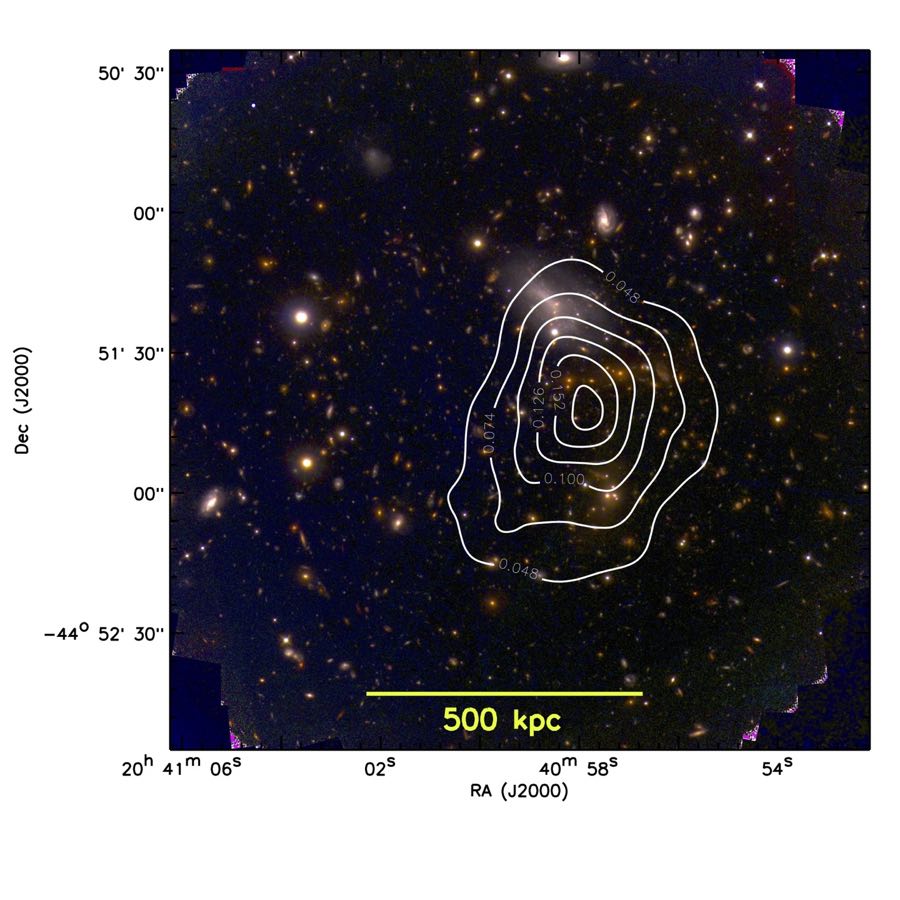}
\caption{Two-dimensional mass distribution of \idcs~(left) and~\spt~(right). The contours are linearly spaced. Because we do not lift the mass-sheet degeneracy, the contour labels do not represent the absolute mass density. In either case, the cluster is unambiguously detected and appears to have a relaxed morphology. The top panel shows only the mass reconstruction whereas in the bottom panel we overlay the mass contours on the color composite images created from the HST imaging data. The ACS/F606W (WFC3/F814W), WFC3/F105W, and WFC3/F140W filers are used to represent the intensities in blue, green, and red for \idcs~(\spt). Our bootstrapping analysis proves that the mass centroids are consistent with the cluster galaxy centroids.
} \label{fig_massmap}
\end{figure*}

Our mass reconstruction results are displayed in Figure~\ref{fig_massmap}. The presence of strong mass concentration is clearly detected in our weak-lensing analysis for both clusters (at the $\mytilde4\sigma$ and $\mytilde6\sigma$ levels for \idcs~ and \spt, respectively). Within the effective smoothing scale $\mytilde30\arcsec$, both clusters appear to possess a relaxed morphology. 

\begin{figure*}
\includegraphics[width=9cm]{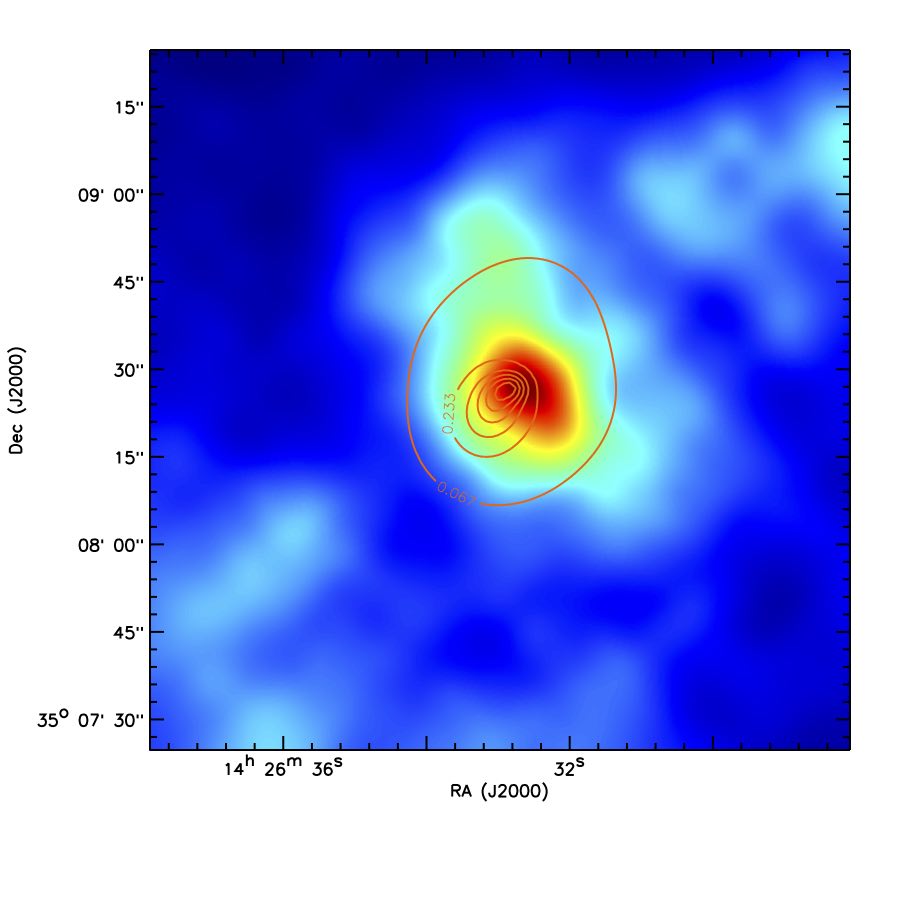}
\includegraphics[width=9cm]{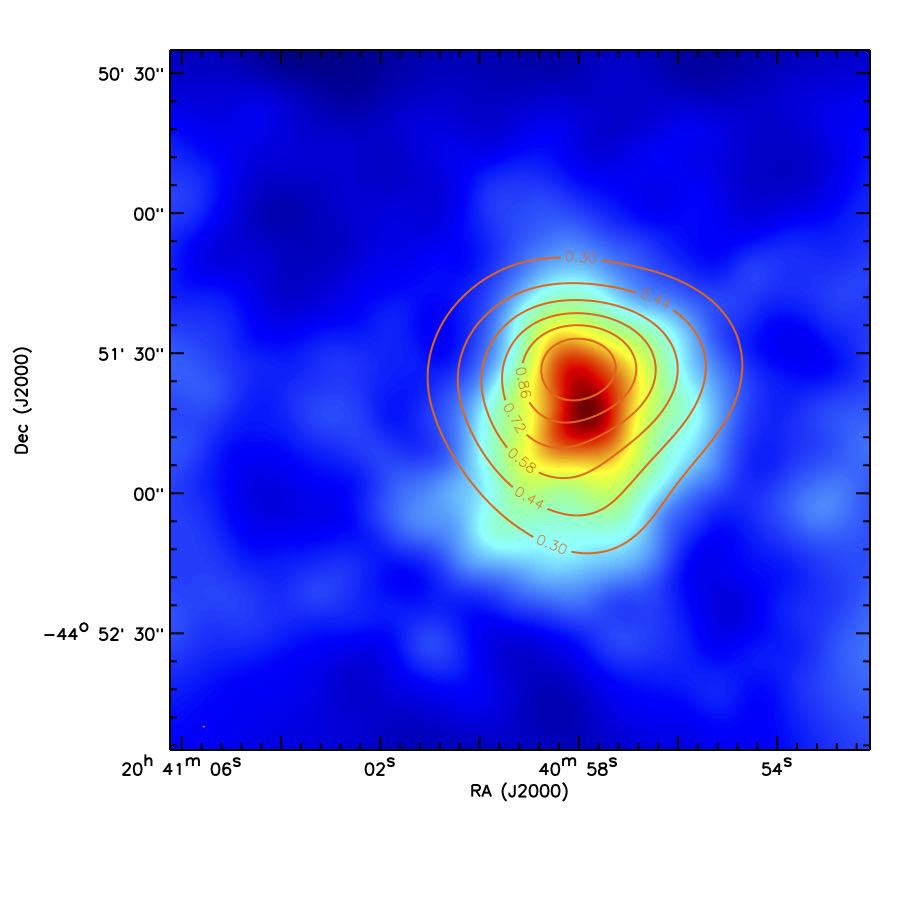}
\caption{Comparison of X-ray emission and mass distribution for 
 \idcs~(left) and~\spt~(right). The contours represent the X-ray intensity obtained from the archival Chandra data after point source removal. We adaptively smooth the X-ray data in such a way that the minimum significance becomes 3$\sigma$.
The contours are spaced linearly. We rescale the X-ray intensity 
to range between zero and unity.
The background is color-coded with the mass density (same as shown in the top panel of Figure~\ref{fig_massmap}). The comparison shows that in both clusters the mass centroid is in good agreement with the X-ray peak. 
} \label{fig_massoverxray}
\end{figure*}

In the bottom panel of Figure~\ref{fig_massmap}, we display the same mass contours on top of the color-composite images. The offsets between the mass peak and the BCG candidate location are $\mytilde4\arcsec$ and $\mytilde11\arcsec$ for \idcs~and~\spt, respectively.
Based on bootstrapping\footnote{We resample shear catalogs with replacement 1000 times. For each realization, we create a mass map and measure the centroid.}, we estimate the uncertainties of the mass centroids to be $\mytilde9\arcsec$ and $\mytilde4\arcsec$ for \idcs~and \spt, respectively (Table 2).
The significance of the BCG-mass peak offset is weak ($\lesssim1\sigma$) 
for \idcs~whereas the offset has a $2.5\sigma$ significance in the \spt~result.
Comparison of centroids between X-ray emission and lensing mass is also a useful diagnostic to infer the clusters' dynamical stage. 
Both clusters have been observed with the {\it Chandra} X-ray telescope. We downloaded the data sets for \idcs~(PI. M. Brodwin) and~\spt~(PI. S. Murray)
from the Chandra Data Archive\footnote{http://cda.harvard.edu/chaser/}. The exposure time for each cluster observation is $\mytilde100$~ks. We identified and removed X-ray point sources using the ``wavdetect" package. The resulting images were adaptively smoothed with a minimum significance of 3$\sigma$. Figure~\ref{fig_massoverxray} displays the X-ray contours obtained in this way overlaid on the mass reconstruction.
For \idcs, the X-ray peak is offset by $\mytilde5\arcsec$ to the east with respect to the lensing peak. This offset is insignificant (only at the $\mytilde0.5\sigma$ level). \spt~possesses its X-ray peak offset to the north by $\mytilde7\arcsec$. With its smaller centroid error of $\mytilde4\arcsec$ considered (Table 2), the offset significance is not strong (at the $\mytilde1.6\sigma$ level). 
In our computation of these significances, we did not consider the centroid errors of the X-ray peaks and believe that doing so will reduce the significances for both clusters. Thus, we conclude that neither cluster shows a large enough offset between X-ray and mass peaks to challenge the validity of our mass estimation utilizing these centers.

\begin{deluxetable}{ccccc}
\tabletypesize{\scriptsize}
\tablecaption{Mass reconstruction centroids and their uncertainties}
\tablenum{2}
\tablehead{\colhead{Galaxy cluster} &  \colhead{R.A.} & \colhead{Decl.} & \colhead{$\sigma_{R.A.}$} & \colhead{$\sigma_{Decl.}$} }
\tablewidth{0pt}
\startdata
\spt & 20:40:57.85 & -44:51:42.4 & 4\farcs0 & 4\farcs4 \\
\idcs & 14:26:32.66 & +35:08:26.9 & 7\farcs5 & 10\farcs7
\enddata
\end{deluxetable}

\subsection{Mass Estimation}
\label{section_mass_estimation}

We determine cluster masses by fitting Navarro-Frenk-White (NFW; Navarro et al. 1997) profiles to reduced tangential shears. Reduced tangential shears are estimated by evaluating
\begin{equation}
 g_T  = -  g_1 \cos 2\phi - g_2 \sin 2\phi \label{tan_shear},
\end{equation}
where $\phi$ is the position angle of the object measured counter-clockwise with respect to the reference axis. One of the important decisions in constructing the profile is the choice of the cluster center because the shape of the azimuthally averaged profile particularly at small radii is sensitive to this choice.
However, as discussed in \textsection\ref{section_mass_reconstruction}, the centroids of the mass, luminosity, and X-ray emission are close to one another in both clusters. Thus, the dependence of the shape of the tangential shear profile on the choice among the three centroids is weak. In this paper, we present the results obtained when the X-ray peak is chosen.

\begin{figure*}
\includegraphics[width=9cm]{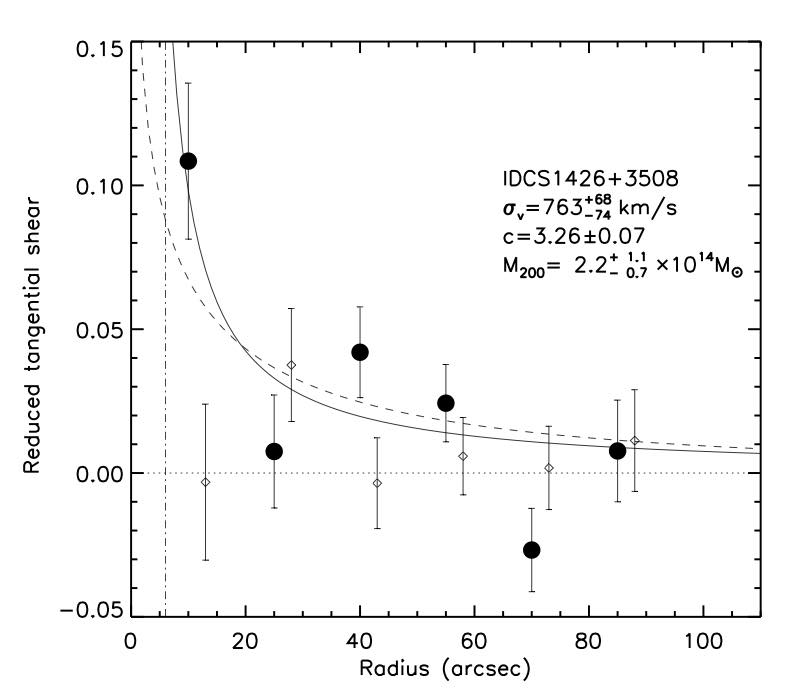}
\includegraphics[width=9cm]{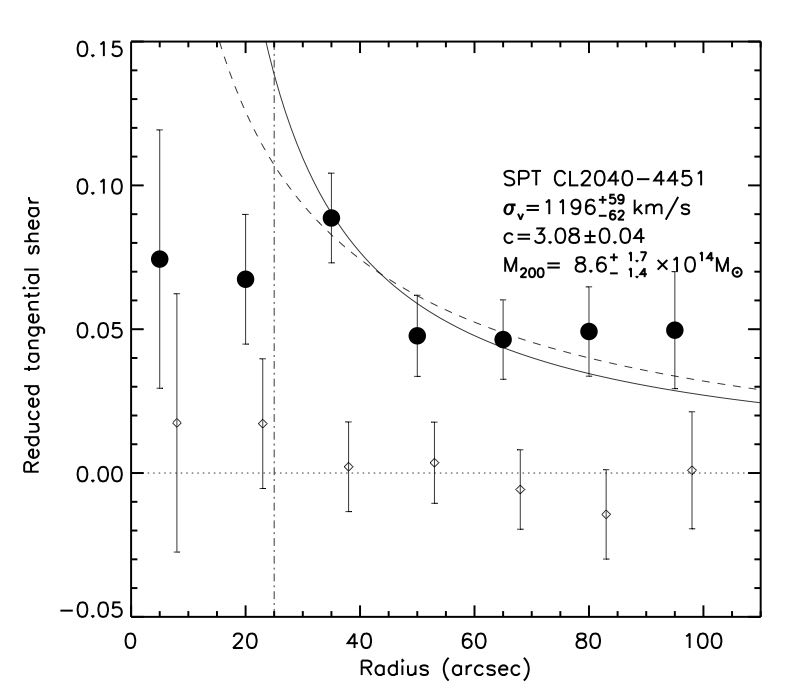}
\caption{Tangential shear profile of \idcs~and~\spt.  The filled circles are tangential shears with the centers fixed on the X-ray peaks. The open diamond symbols are computed in the same way except that we rotate the source galaxies by 45\degr. These so-called B-mode statistics are useful to test the level of systematics in weak-lensing studies. As shown here, the B-mode signals are consistent with zero in both clusters. The dot-dashed lines depict limits inside which we exclude data for SIS and NFW fitting.
The velocity dispersion $\sigma_v$ is computed based on SIS fitting results. For NFW fitting, we assume the mass-concentration relation of Dutton \& Maccio (2014) and obtain best-fit concentration parameters. The displayed masses are derived from this NFW fitting.
} \label{fig_t_shear}
\end{figure*}

Figure~\ref{fig_t_shear} displays the reduced tangential shear profiles of~\idcs~and~\spt. Lensing signals are clearly seen in both clusters. The filled circles are tangential shears with the centers fixed on the X-ray peaks. The open diamond symbols are computed in the same way except that we rotate the source galaxies by 45\degr. These so-called B-mode statistics are useful to test the level of systematics in weak-lensing studies. As shown here, the B-mode signals are consistent with zero in both clusters.

Together with the choice on the cluster center, another important question is where to put the lower limit radius $r_{min}$, inside which we exclude the signal in model fitting. There are many reasons why we avoid using the signals from the very central region of clusters. First, the signal amplitude there is sensitive to the choice of the cluster center and the presence of any substructure. Second, we expect a high cluster member contamination rate. Third, current theories do not agree on the exact behavior of the inner cluster density profile. Fourth, in massive clusters our weak-lensing assumptions break down near and inside the Einstein radius. 

No consensus has been reached on the optimal strategy regarding the evaluation of $r_{min}$, although we understand that the choice should depend on the cluster mass because it determines the size of the cluster core, where the aforementioned issues become important. In this study, we use an Einstein radius from our singular isothermal sphere (SIS) fitting as a guideline to determine the cutoff value. We begin by using all data points available (i.e., $r_{min}=0$) and fitting an SIS model to them. We choose $r_{min}$ to be four times the resulting Einstein radius. Then, we apply this cut to the tangential shear profile and iterate the procedure until the result converges.

The dot-dashed lines (Figure~\ref{fig_t_shear}) depict the $r_{min}$ values ($\mytilde6 \arcsec$ and $\mytilde25\arcsec$ for \idcs~and~\spt, respectively), inside which we exclude data for SIS and NFW fitting. 
The velocity dispersion $\sigma_v$ is computed based on SIS fitting results. For NFW fitting, we assume the mass-concentration relation of Dutton \& Maccio (2014) and obtain best-fit concentration parameters. Dutton \& Maccio (2014) update the mass-concentration relation of previous studies (e.g., Duffy et al. 2008) using
the cosmological parameters favored by Planck collaboration et al. (2016). We will further discuss the impact of different mass-concentration relations on cluster masses in \textsection\ref{section_mcmc}.
The resulting masses of \idcs~and~\spt~are
$M_{200}=2.2_{-0.7}^{+1.1}\times10^{14}~M_{\sun}$ and $8.6_{-1.4}^{+1.7}\times10^{14}~M_{\sun}$, respectively. For \idcs, the $r_{min}$ value is negligibly small and thus does not affect the cluster mass. However, if we did not exclude the inner data points, the $M_{200}$ value of \spt~would be $\mytilde25$\% smaller.

\section{DISCUSSIONS}
\label{section_discussion}
\subsection{Non-statistical Uncertainties in Mass Estimation}
The error bars of the cluster masses in \textsection\ref{section_mass_estimation} include only statistical errors due to the finite number of sources. Control of systematic errors is the theme of the next-generation cosmology surveys because the interpretation is unlikely to be limited by statistical uncertainties. Here, we discuss non-statistical sources of errors that potentially affect our mass estimation. These additional errors should be considered when we interpret the cosmological significance of the two clusters presented in this paper.

\subsubsection{Source Redshift Uncertainty}
As stated in \textsection\ref{section_source_selection}, we use the UVUDF photometric redshift catalog of Rafelski et al. (2015) to infer the redshift distribution of the sources in our cluster fields. The procedure leads to non-negligible errors in cluster mass estimation, which can be categorized into three types: 1) systematic errors in UVUDF, 2) sample variance in UVUDF (i.e., UVUDF does not represent the mean), and 3) sample variance in the cluster fields (i.e., the cluster field's statistics depart from the mean). 

As for the first issue, Rafelski et al. (2015) compare their photometric redshifts with spectroscopic redshifts and find that both the scatter and catastrophic outlier fraction are small. With respect to the previous UDF photometric redshift of Coe et al. (2006), the normalized median absolute deviation decreases by more than a factor of two. Also, the outlier fraction is only $\mytilde2.4$\% (the value from the Coe et al. 2006 result is $\mytilde16.4$\%). The majority of the spectroscopic sources are relatively bright and at $z\lesssim2$ (the maximum spectroscopic redshift is $z\sim6$). Therefore, one should be careful in extending this comparison result to fainter sources. The solid confirmation of the accuracy in this regime is not feasible until future telescopes with extremely large apertures become available. Nevertheless, we do not suspect that there is any surprisingly large departure for non-spectroscopically confirmed fainter sources because the extreme depth of UVUDF (in most filters the 5$\sigma$ limiting magnitude is $\mytilde30$) provides high-quality photometry of these ``faint" populations. The multi-wavelength observations enable robust detection of the Lyman and Balmer breaks at $0.8\lesssim z \lesssim 3.4$. 

The second and third issues are closely related in the sense that both arise from the sample variance. However, it is often useful to distinguish them. For example in the case that the second issue is negligible, the third issue only introduces scatters (statistical errors) for individual clusters without biasing their average. In Jee et al. (2014b), we investigated the second issue in our study of the ``El Gordo" cluster utilizing the photometric redshift catalogs of GOODS-N and GOODS-S (Dahlen et al. 2010). Since UDF is a subfield of GOODS-S, we first checked how well the small $\mytilde10~\mbox{arcmin}^{2}$ UDF result agreed with the result from the entire $\mytilde160~\mbox{arcmin}^{2}$ GOODS-S field. The difference was only $\mytilde0.005$ in terms of $\beta$. This implies that the sample variance might be small at least within the $\mytilde160~\mbox{arcmin}^2$ field. To estimate the Poisson scatter, we defined eight non-overlapping UDF-size regions within the GOODS-S field and obtained a standard deviation of $\sigma_{\beta}=0.005$, which translates to $\mytilde2$\% in the mass uncertainty of ``El Gordo." This level of uncertainty is much smaller than the statistical one. We repeated the experiment within the GOODS-N field and found that the mean value of $\beta$ there was higher than the southern field value by $\Delta \beta=0.006$ while the scatter is $\sigma_{\beta}=0.008$ from seven UDF-size subfields within GOODS-N. Because the two GOODS fields are widely separated, we concluded in Jee et al. (2014b) that the source redshift uncertainty is not a dominant source of error in mass estimation of ``El Gordo."

\begin{figure}
\centering
\includegraphics[width=8.8cm]{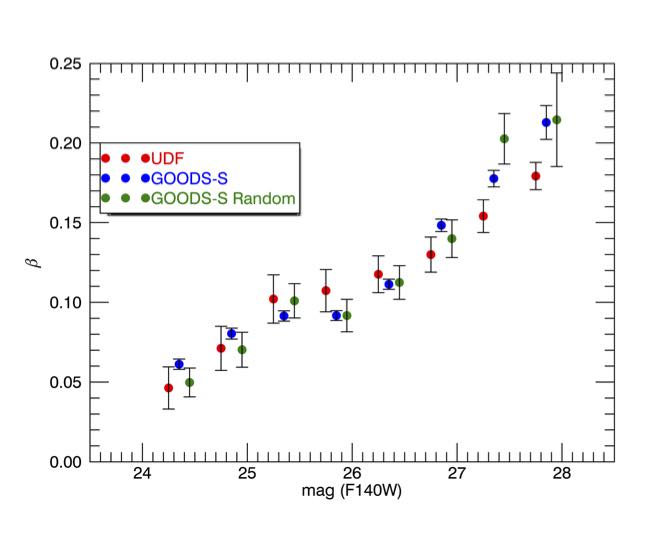}
\caption{Redshift estimation of source population. We show the \spt~ result here. We compare the $\beta$ values as a function of F140W magnitude between UDF and GOODS-S. Because the GOODS-S field does not have an F140W filter coverage, we performed photometric transformation using F105W and F160W. We find that not only the global mean value of $\beta$ is in good agreement, but also the magnitude-dependent trend is similar. The green symbol represents the result when we randomly select a UDF-size field within the GOODS-S field. We do not observe any significant departure from the results obtained within the entire GOODS-S field or UDF. We offset the blue and green symbols slightly to avoid clutter.
}
\label{fig_beta}
\end{figure}

In this paper, we repeat the experiment of Jee et al. (2014b). However, we use only the GOODS-S field, which has been covered by the WFC3-IR filters in the CANDELS program. Because the photometric redshift catalog of the CANDELS program is not in the public domain yet whereas the multi-wavelength catalog is publicly available (Guo et al. 2013), we cross-match the old photo-$z$ catalog of Dahlen et al. (2010) with the Guo et al. (2013) photometry catalog. In Figure~\ref{fig_beta}, we compare the $\beta$ values as a function of F140W magnitude between GOODS-S and UDF when we apply the same source selection criteria; we only display the result for the case of \spt~ here.
Because the GOODS-S field does not have an F140W filter coverage, we performed
photometric transformation to estimate the F140W magnitude using the GOODS F105W and F160W photometry. The $\beta$ value in each magnitude bin is in good agreement between the two fields, supporting our previous claim in Jee et al. (2014b). Considering the UDF field is smaller than the GOODS-S field by more than an order of magnitude, the good agreement supports our previous claim that the sample variance does not vary significantly within a $\mytilde160~\mbox{arcmin}^{2}$ field. 

For \spt, the global $\beta$ average from GOODS-S is $\left< \beta \right>=0.122$, which agrees well with the UDF estimate $\left< \beta \right>=0.120$. The green filled circles represent the result when we randomly select one UDF-size field within the GOODS-S field. Apart from the enlarged error bars due to reduced source numbers, the trend is consistent with those from UDF or the entire GOODS-S field; the average $\beta$ value from this sub-sample is $\left< \beta \right>=0.125$
When we select eight non-overlapping UDF-size areas within the GOODS-S field and repeat the estimate, we obtain a standard deviation of $\sigma_{\beta}=0.004$, which again indicates that the sample variance is sub-dominant within the GOODS-S field. We observe a consistent trend for \idcs.

Therefore, from the current test with the GOODS-S field and our previous experiment in Jee at al. (2014b), we conclude that the effect of the sample variance on the cluster mass uncertainty is smaller in this study than the case for ``El Gordo" because the current imaging data is much deeper (thus more cosmic volume along the line-of-sight direction). If we adopt $\sigma_{\beta}=0.008$ (the difference in $\beta$ between the GOODS-N and -S in Jee et al. 2014b) as the uncertainty of our $\beta$ estimation, the resulting mass uncertainty from this source becomes $\mytilde6$\% and $\mytilde9$\% for \spt~ and \idcs, respectively.

\subsubsection{Shear Calibration}
\label{section_shear_calibration}
Our weak-lensing pipeline has been applied to a wide range of both space- and ground-based data (e.g., Jee et al. 2009; 2011; 2013; 2016). We have found that the dominant shear bias comes from the multiplicative factor mentioned in \textsection\ref{section_shape_measurement}, which can be calibrated out using image simulations. In this paper, however, we choose to perform the calibration utilizing the existing ACS imaging data that overlaps some of the ``See Change" WFC3-IR fields.
This is because a high-fidelity end-to-end simulation is not yet feasible for the WFC3-IR instrument, whose understanding is still growing. 

We use the following three ``See Change" cluster fields: SPT-CL J2106-5844 (Foley et al. 2011), SPT-CL0205-5829 (Stalder et al. 2013), and \idcs. The last cluster is of course one of the two clusters studied here. We processed both ACS F606W and WFC3-IR F140W images of the three clusters using our weak-lensing pipeline and found a total of $\mytilde5,200$ common objects. Out of these common objects, we discarded $\mytilde48$\% that did not meet our source selection criteria (\textsection\ref{section_source_selection}).
We compared the $e_1$ and $e_2$ components of the remaining $\mytilde2700$ objects between the ACS and WFC3 and derived the multiplicative factor. 
We found that the multiplicative factor agrees within the 1\% level between the two $e_1$ and $e_2$ ellipticity components. In addition, the field-to-field (i.e., cluster-to-cluster) variation is small ($\lesssim2$\%).

\begin{figure}
\centering
\includegraphics[width=8.8cm]{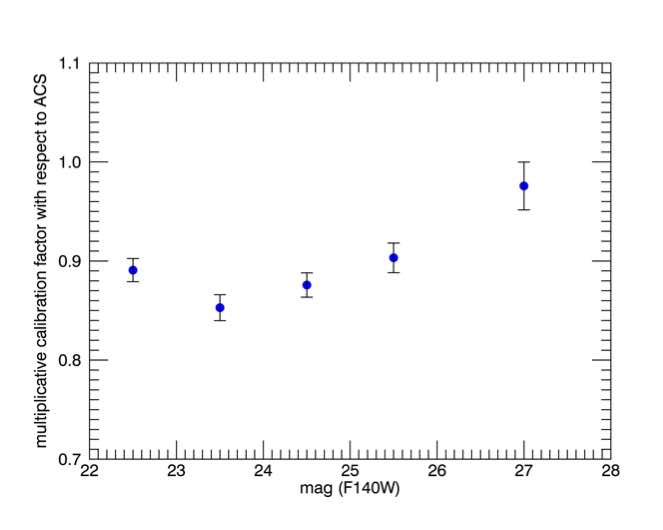}
\caption{Shear calibration with respect to ACS shapes. This calibration is performed with a total of $\mytilde2700$ common objects between ACS and WFC3 images. Overall, the ellipticities measured from WFC3 images are systematically lower. When we apply this calibration to our cluster weak-lensing analysis, we derive a global calibration factor after taking into account both the magnitude dependence and the source magnitude distribution.}
\label{fig_shear_calib}
\end{figure}

Figure~\ref{fig_shear_calib} shows the shear calibration of WFC3 ellipticities with respect to ACS values as a function of the F140W magnitude.
For the entire source population, we determine the average calibration factor to be 1.11 (we need to multiply this calibration factor to match the ACS results). However, we
caution readers that the exact number will depend on the details of one's pipeline implementation and the current result may not be directly applicable to other pipelines.
Nevertheless, we note that the ellipticities from the WFC3 IR data are $\mytilde10$\% smaller on average in our case. 
The precise cause of this large dilution is unknown. However, we speculate that a significant contributor may be the aliasing effect due to the PSF undersampling in the WFC3-IR detector (\textsection\ref{section_undersampling}). 

A self-consistency test is carried out by applying the shear calibration obtained from the above ellipticity-to-ellipticity comparison to weak-lensing analysis. The cluster mass derived from the ACS and WFC3 shapes is in excellent agreement, although we select different galaxies for our lensing sources, for which we also separately estimate their redshift distributions. The weak-lensing studies for SPT-CLJ2106-5844 and SPT-CL0205-5829 will be presented in separate publications.
From bootstrap analysis, we derive a shear calibration uncertainty, which translates to a cluster mass uncertainty of $\mytilde4$\%.

\subsubsection{Mass-Concentration Relation Assumption}
\label{section_mcmc}
We assume the Dutton \& Maccio (2014) mass-concentration relation in \textsection\ref{section_mass_estimation}. This mass-concentration assumption is needed because the two parameters of an NFW profile are highly degenerate and cannot be constrained simultaneously given the noise level of the data. 

It is important to remember that the mass-concentration relation depends on the assumed cosmological parameters (e.g., Dutton \& Maccio 2014) and the choice has a non-negligible impact on the mass estimation.
For example, if we assumed the Duffy et al. (2008) relation based on the WMAP5 cosmology instead, the mass of \spt~would increase by $\mytilde25$\% favoring a low concentration $c\sim2.3$; the reduced $\chi^2$ values in both cases are very similar ($\chi_{\nu}\sim0.52$).

Another important issue is the uncertainty of the behavior of the mass-concentration relation for massive high-redshift clusters because they are also rare in numerical simulations. Neither Duffy et al. (2008) nor Dutton \& Maccio (2014) constrain 
the relation beyond $M_{200}\gtrsim10^{14}M_{\sun}$ in the redshift regime of the
clusters studied here.
If the true mass-concentration relation departs from the extrapolation of the lower-mass cluster results, it is obvious that using the above relations biases cluster masses. 
Finally, since the mass-concentration relation is only the mean property derived from a population with considerable scatter, individual cluster mass estimates based on the mean relation are subject to additional uncertainties. In particular, when one studies an extremely rare cluster such as \spt, it is not easy to justify the use of the mean relation in the lower-mass regime in deriving the mass of the exceptional system.

We study the impact of this mass-concentration scatter using Markov Chain Monte Carlo (MCMC) analysis by marginalizing over ranges of concentration $c$ and scale radius $r_s$ parameters. The centers of the concentration intervals are chosen to be the converged values with the Dutton \& Maccio (2014) relation presented in \textsection\ref{section_mass_estimation}. We set the width of the $\Delta c$ interval to 1 (i.e., $[2.08,4.08]$ and $[2.26,4.26]$ for \spt~and \idcs, respectively). The mass
range is wide, from $\mytilde10^{13}~ M_{\sun}$ to 
$\mytilde 10^{16}~M_{\sun}$ in Dutton \& Maccio (2014).
The interval in scale radius is determined by finding the corresponding values of $r_s$ for the upper and lower limits of $c$ while the same mass-concentration relation is assumed.
Out of 60,000 MCMC chains, we obtain $M_{200}=8.2\pm1.9\times10^{14} M_{\sun}$
and $1.8\pm0.9 \times10^{14} M_{\sun}$ for \spt~and~\idcs, respectively. These results are consistent with the values determined by assuming the mass-concentration relation of Dutton \& Maccio (2014). We cannot guarantee that the prior ranges adopted above span the most probable parameter space sufficiently. However, we believe the experiment explores the possible range of the masses when the true relation moderately scatters around the Dutton \& Maccio (2014) relation.

\subsubsection{Large-scale Structure, Triaxiality, and Departure from NFW}

Because lensing measures projected mass distributions while
what we estimate are de-projected masses of clusters, a few issues arise. Here we discuss the effects of the large-scale structure, triaxiality, and departure of cluster profile from NFW.

Background large-scale structures uncorrelated with clusters  contaminate the lensing signal from the galaxy clusters that we study. Hoekstra (2003) estimated the effect quantitatively by integrating a nonlinear power spectrum along the line-of-sight direction and concluded that the contamination is one of the limiting factors in accurate cluster mass determination.

Following the method of Hoekstra (2003), we estimate that the uncorrelated large-scale structure provides additional correlated noise of $\sigma_{\gamma}\sim0.01$ within the angular scale measured in this study. Our estimate is about a factor of two larger than the value presented in Hoekstra (2003) because the mean redshift of the source is much higher in our data. This will increase the mass error bars presented in \textsection\ref{section_mass_estimation} by $\mytilde22$\% and $\mytilde17$\% for \spt~and~\idcs, respectively. 

We assume a spherical NFW profile when we estimate the mass, although real clusters deviate from this assumption, having substructures, triaxiality, etc.
These issues have been studied through numerical simulations
 (e.g., Meneghetti et al. 2010; Becker
\& Kravtsov 2011; Oguri \& Hamana 2011).
Becker \& Kravtsov (2011) found that the intrinsic
scatter is $\mytilde20$\% for massive halos. And more recently, Gruen et al. (2015) used a semi-analytical model and also showed that neglecting intrinsic profile variations causes significant underestimation of cluster mass uncertainties. Therefore, one should include the impact of this intrinsic cluster profile variation together with the above sources of error when discussing cosmological implications for the existence of our clusters.

\subsection{Rarity}
Currently, \spt~is the most massive cluster at $z\gtrsim1.5$ confirmed by weak lensing.
According to our hierarchical structure formation paradigm, massive high-redshift galaxy clusters are rare. 
We will present an extensive analysis of the topic using all available massive high-$z$ clusters from the current ``See Change" project and previous archival programs in future publications. Hence, the scope of the current paper is to examine the rarity of the two clusters studied here using a traditional mass function method.

We adopt the mass function of Tinker et al. (2008) to estimate our cluster abundance using the best-fit cosmological parameters published in Planck Collaboration et al. (2016). The estimated abundances of \idcs~ and \spt~ are $\mytilde1200$ and $\mytilde1$, respectively, over the full sky if we choose the central values of our mass estimates.
The weak-lensing mass implies that \idcs~is not an exceptionally rare object like \spt. Neverthless, observational difficulties will continue to hamper us from dramatically increasing the number of known clusters at such high redshifts.
Our weak-lensing mass estimation of \spt~confirms that the cluster is a rare object as first mentioned by Bayliss et al. (2014).
Given the total 2500 sq. deg. area of the SPT-SZ survey (Bleem et al. 2015), the abundance quoted above for the full sky should decrease by a factor of $\mytilde16$. The resulting abundance $\mytilde0.07$ suggests that perhaps the survey had to be very ``lucky" to find a such a rare cluster. If we choose the initial survey area 720 sq. deg. where \spt~was discovered (Bayliss et al. 2014), the
expected abundance within that survey area becomes $\mytilde2\times10^{-2}$.
A conservative lower-limit of our mass estimate may be the MCMC result in \textsection\ref{section_mcmc}, where we marginalize over ranges of concentration values and scale radii. With this choice of threshold $\mytilde6\times10^{14}~M_{\sun}$, the expected abundance becomes $\mytilde0.3$ within the initial 720 deg$^2$ survey area. 

It is premature to argue that the existence of \spt~yields a significant tension to the standard $\Lambda$CDM model of our universe; Williamson
et al. (2011) claimed that none of the most massive SPT clusters are individually in significant tension 
with the $\Lambda$CDM cosmological model. However, it is still interesting to note that recent studies have found a number of massive high-redshift galaxy clusters whose masses are not comfortably reconciled with the predictions of
the current $\Lambda$CDM models when their parent survey volumes are chosen as the normalization volume (e.g., Jee et al. 2009, Menanteau et al. 2012; Foley et al. 2011; J. Kim et al. in prep). For example, 
a contender at a similarly high redshift is XMMU J2235.3-2557 at $z=1.4$, whose virial mass from both weak-lensing (Jee et al. 2009) and X-ray (Rosati et al. 2009) exceeds $\mytilde8\times10^{14}~M_{\sun}$. 
Because of many subtleties, interpretation is difficult when one bases his/her analysis on a single cluster detection (e.g., Davis et al. 2011; Waizmann et al. 2011; Hotchkiss 2011).
In addition, these extremely massive clusters are also rare in numerical simulation results, and thus the behavior of the cluster mass function and the mass-concentration relation in this extreme regime is still uncertain.
More statistically robust studies will become possible after high-redshift cluster mass functions are constructed from a larger sample.

\subsection{Comparison with Other Studies}

The first mass measurement of \spt~was presented in Reichardt et al. (2013), who quoted $M_{500}=(3.21\pm0.79)\times10^{14}M_{\sun}$ based on the initial 720 sq. degree SZ data. Assuming the NFW profile with a mass-concentration relation of Duffy et al. (2008), Bayliss et al. (2014) extrapolated this $M_{500}$ mass to $M_{200}=(5.8\pm1.4)\times10^{14}M_{\sun}$. This SZ mass is marginally
consistent with our weak-lensing mass\footnote{Perhaps, the difference decreases slightly if we take into account the fact that the cluster redshift is assumed to be $z_{phot}=1.35$ instead of $z_{spec}=1.48$ in Reichardt et al. (2013).}.

The velocity-dispersion estimate by Bayliss et al. (2014) is highly uncertain because the number of available member galaxies is small (15) and the candidate selection is biased toward ``blue" galaxies. Nevertheless, their value $\sigma_{v,gap}=1500\pm520~\mbox{km~s}^{-1}$ obtained using the gapper statistic is marginally consistent with 
our weak-lensing estimate $\sigma_{v,WL}=1196_{-62}^{+59}\mbox{km~s}^{-1}$, which is computed from the SIS fitting result. As Bayliss et al. (2014) discussed,  velocity dispersions are likely to be inflated when star-forming galaxies (thus infalling to the cluster) are used. 

\idcs~was quoted by Brodwin et al. (2016) as the most massive galaxy cluster at $z>1.5$. Using their 100~ks Chandra observation and the various scaling relations published in the literature, they measured the cluster mass to span the range $M_{500}=2.3-3.3\times10^{14}M_{\sun}$. This result is in good agreement with their SZ mass of $M_{500,Y_X}=2.6_{-0.5}^{+2.6}\times10^{14}M_{\sun}$. We obtained the first weak-lensing mass $M_{200}=2.3_{-1.4}^{+2.1}\times10^{14}M_{\sun}$ of~\idcs~using the ACS observation (Mo et al. 2016). This value is
approximately at the lower end of the other results, yet still with overlapping error bars when the other results are rescaled to the values at $r_{200}$.
For example, the SZ mass extrapolated to $M_{200}$ is $4.1\pm1.1\times10^{14}M_{\sun}$ (Brodwin et al. 2012).

The current weak-lensing mass estimate $M_{200}=2.2_{-0.7}^{+1.1}\times10^{14}M_{\sun}$ from our new HST imaging data agrees well with the result of Mo et al. (2016). Although the much higher source density from the current WFC3 imaging data makes the statistical
uncertainty ∼40\% smaller, our reduced error bars still overlap those of the SZ-based results.

\begin{figure}
\centering
\includegraphics[width=8cm]{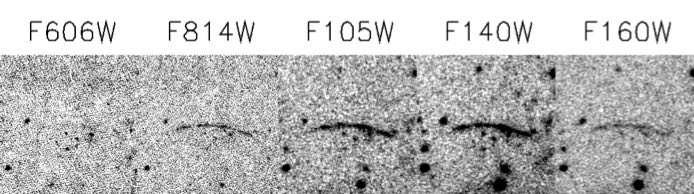}
\caption{Multiband observation of the giant arc in \idcs. The giant arc is clearly a ``dropout" in the very deep F606W image whose exposure time is 21,760s. }
\label{fig_arc}
\end{figure}

A strong-lensing analysis of \idcs~was presented by Gonzalez et al. (2012) using the prominent arc $\mytilde15\arcsec$ north of the BCG. Without the knowledge of the redshift of the arc, they studied a plausible range of the extrapolated mass at $r_{200}$. Although the spectroscopic redshift of the giant arc is still unknown to date, it is possible to constrain the range based on our broadband photometry.
As mentioned by Brodwin et al. (2016), this arc is a dropout in the deep (21,760 s, 28th mag, $10\sigma$) F606W image (Figure~\ref{fig_arc}).
Adopting the redshift
interval $4.5\lesssim z \lesssim 6$ of Brodwin et al. (2016)
and following the extrapolation scheme of Gonzalez et al. (2012), we estimate that the extrapolated mass ($M_{200}$) ranges from $\mytilde2\times10^{14}M_{\sun}$ to $\mytilde3\times10^{14}M_{\sun}$. This range
nicely brackets the central value of our weak-lensing result.

\section{CONCLUSIONS}
\label{section_summary}
We have presented weak-lensing analysis of two very distant galaxy clusters \idcs~and~\spt~at $z=1.75$ and $1.48$, respectively. This is the first weak-lensing study from the See Change project, which targets 12 massive cosmologically interesting high-$z$ clusters with HST/WFC3. 

Weak-lensing signals are clearly detected in both clusters. 
The reconstructed mass map of each cluster shows a single mass peak, whose centroid agrees with the cluster galaxies.
We estimate the mass of \idcs~to be $2.2_{-0.7}^{+1.1}\times10^{14}M_{\sun}$, which agrees with our previous weak-lensing mass based on ACS imaging data. This mass is also consistent with the value extrapolated from the strong-lensing, X-ray, and SZ studies.

We find that the mass of \spt~is extremely high ($8.6_{-1.4}^{+1.7}\times10^{14} M_{\sun}$) for a cluster at such a high redshift ($z=1.48$). This mass is also consistent with the extrapolated mass from the SZ study.
Adopting the central value of our weak-lensing result, the expected abundance of such massive clusters is only$\mytilde0.07$  in the parent 2500 sq. deg. survey. Although it is premature to argue that the discovery of \spt~yields a tension with the current $\Lambda$CDM cosmology,
we note that similarly rare massive high-redshift clusters have been occasionally reported in various surveys. We believe that it is worth performing a more statistically robust analysis with the combination of other massive high-redshift clusters after marginalizing over theoretical uncertainties such as the mass function and mass-concentration relation.

In addition to shot noise, we have investigated the impact of the sample variance in the source galaxy redshift estimation and the uncertainty in the mass-concentration relation on the cluster mass error. We find that, although the impact of the sample variance is non-negligible, more important is the assumption on the galaxy cluster mass profile. Our experiment shows that different assumptions on the mass-concentration relation affect the cluster mass by $\mytilde20$\%.

This is the first weak-lensing study using the WFC3. We have demonstrated that, despite the slightly larger PSF than that of the ACS, the higher sensitivity provides a greater leverage for weak lensing, when we study high-$z$ ($z\gtrsim1$) clusters, where the lensing signal mostly comes from very distant, faint galaxies. 

\acknowledgments
{We thank the anonymous referee for carefully reading our manuscript and providing valuable comments, which improved the quality of the paper.
We also thank Douglas Applegate, Lindsey Bleem, Brian Hayden, David Rubin, Henk Hoekstra, and Tim Schrabback for useful discussions.

Support for the current HST program was provided by
NASA through a grant from the Space Telescope Science
Institute, which is operated by the Association of Universities
for Research in Astronomy, Incorporated, under NASA contract
NAS5-26555. 

M.J.J. acknowledges support for the current research 
from the Korea Astronomy and Space Science Institute under the R\&D
program supervised by the Ministry of Science, ICT and Future Planning and the National Research Foundation of Korea under the program 2017R1A2B2004644.}


\begin{thebibliography}{}

\bibitem[Albrecht et al.(2006)]{2006astro.ph..9591A} Albrecht, A., Bernstein, G., Cahn, R., et al.\ 2006, arXiv:astro-ph/0609591 
\bibitem[Applegate et al.(2014)]{2014MNRAS.439...48A} Applegate, D.~E., von der Linden, A., Kelly, P.~L., et al.\ 2014, \mnras, 439, 48 
\bibitem[Bayliss et al.(2014)]{2014ApJ...794...12B} Bayliss, M.~B., Ashby, M.~L.~N., Ruel, J., et al.\ 2014, \apj, 794, 12 
\bibitem[Bajaj (2016)]{2016ISR} Bajaj, V. 2016, Space Telescope WFC3 Instrument Science Report 2016-02
\bibitem[Becker \& Kravtsov(2011)]{2011ApJ...740...25B} Becker, M.~R., \& Kravtsov, A.~V.\ 2011, \apj, 740, 25 
\bibitem[Beckwith et al.(2006)]{2006AJ....132.1729B} Beckwith, S.~V.~W., Stiavelli, M., Koekemoer, A.~M., et al.\ 2006, \aj, 132, 1729 
\bibitem[Bleem et al.(2015)]{2015ApJS..216...27B} Bleem, L.~E., Stalder, B., de Haan, T., et al.\ 2015, \apjs, 216, 27 
\bibitem[Broadhurst et al.(2005)]{2005ApJ...619L.143B} Broadhurst, T., Takada, M., Umetsu, K., et al.\ 2005, \apjl, 619, L143 
\bibitem[Brown et al.(2006)]{2006PASP..118.1443B} Brown, M., Schubnell, M., \& Tarl{\'e}, G.\ 2006, \pasp, 118, 1443 
\bibitem[Brodwin et al.(2012)]{2012ApJ...753..162B} Brodwin, M., Gonzalez, A.~H., Stanford, S.~A., et al.\ 2012, \apj, 753, 162 
\bibitem[Brodwin et al.(2016)]{2016ApJ...817..122B} Brodwin, M., McDonald, M., Gonzalez, A.~H., et al.\ 2016, \apj, 817, 122 
\bibitem[Davis et al.(2011)]{2011MNRAS.413.2087D} Davis, O., Devriendt, J., Colombi, S., Silk, J., \& Pichon, C.\ 2011, \mnras, 413, 2087 
\bibitem[Dressel et al. (2016)]{2016STSCI} Dressel, L., et al. \ 2016, "Wide Field Camera 3 Instrument Handbook, Version 8.0” (Baltimore: STScI)
\bibitem[Duffy et al.(2008)]{2008MNRAS.390L..64D} Duffy, A.~R., Schaye, J.,
Kay, S.~T., \& Dalla Vecchia, C.\ 2008, \mnras, 390, L64
\bibitem[Ellis et al.(2013)]{2013ApJ...763L...7E} Ellis, R.~S., McLure, R.~J., Dunlop, J.~S., et al.\ 2013, \apjl, 763, L7 
\bibitem[Foley et al.(2011)]{2011ApJ...731...86F} Foley, R.~J., Andersson, K., Bazin, G., et al.\ 2011, \apj, 731, 86 
\bibitem[Gonzalez et al.(2012)]{2012ApJ...753..163G} Gonzalez, A.~H., Stanford, S.~A., Brodwin, M., et al.\ 2012, \apj, 753, 163 
\bibitem[Gruen et al.(2015)]{2015MNRAS.449.4264G} Gruen, D., Seitz, S., Becker, M.~R., Friedrich, O., \& Mana, A.\ 2015, \mnras, 449, 4264 
\bibitem[Guo et al.(2013)]{2013ApJS..207...24G} Guo, Y., Ferguson, H.~C., Giavalisco, M., et al.\ 2013, \apjs, 207, 24 
\bibitem[Hilbert \& McCullough(2011)]{2011ISR}
Hilbert, B. \& McCullough, P.\ 2011, STScI Instrument Science Report: 2011-10
\bibitem[Hoekstra(2003)]{2003MNRAS.339.1155H} Hoekstra, H.\ 2003, \mnras, 339, 1155 
\bibitem[Hotchkiss(2011)]{2011JCAP...07..004H} Hotchkiss, S.\ 2011, JCAP, 7, 004 
\bibitem[Jee et al.(2005)]{2005ApJ...618...46J} Jee, M.~J., White, R.~L., Ben{\'{\i}}tez, N., et al.\ 2005, \apj, 618, 46 
\bibitem[Jee et al.(2007)]{2007PASP..119.1403J} Jee, M.~J., Blakeslee,
J.~P., Sirianni, M., Martel, A.~R., White, R.~L., \& Ford, H.~C.\ 2007a, \pasp, 119, 1403
\bibitem[Jee et al.(2007)]{2007ApJ...661..728J} Jee, M.~J., et al.\ 2007b,
\apj, 661, 728
\bibitem[Jee et al.(2009)]{2009ApJ...704..672J} Jee, M.~J., Rosati, P., Ford, H.~C., et al.\ 2009, \apj, 704, 672 
\bibitem[Jee et al.(2011)]{2011ApJ...737...59J} Jee, M.~J., Dawson, K.~S., Hoekstra, H., et al.\ 2011, \apj, 737, 59 
\bibitem[Jee et al.(2014)]{2014ApJ...783...78J} Jee, M.~J., Hoekstra, H., Mahdavi, A., \& Babul, A.\ 2014a, \apj, 783, 78 
\bibitem[Jee et al.(2014)]{2014ApJ...785...20J} Jee, M.~J., Hughes, J.~P., Menanteau, F., et al.\ 2014b, \apj, 785, 20 
\bibitem[Kozhurina-Platais \& Petro(2012)]{2012wfc..rept....3K} Kozhurina-Platais, V., \& Petro, L.\ 2012, Space Telescope WFC3 Instrument Science Report  
\bibitem[Long et al.(2012)]{2012SPIE.8442E..1WL} Long, K.~S., Baggett, S.~M., MacKenty, J.~W., \& Riess, A.~G.\ 2012, \procspie, 8442, 84421W 
\bibitem[Medezinski et al.(2017)]{2017arXiv170600427M} Medezinski, E., Oguri, M., Nishizawa, A.~J., et al.\ 2017, arXiv:1706.00427 
\bibitem[Melchior et al.(2017)]{2017MNRAS.469.4899M} Melchior, P., Gruen, D., McClintock, T., et al.\ 2017, \mnras, 469, 4899 
\bibitem[Menanteau et al.(2012)]{2012ApJ...748....7M} Menanteau, F., Hughes, J.~P., Sif{\'o}n, C., et al.\ 2012, \apj, 748, 7 
\bibitem[Meneghetti et al.(2010)]{2010A&A...514A..93M} Meneghetti, M., Rasia, E., Merten, J., et al.\ 2010, \aap, 514, A93 
\bibitem[McCullough(2008)]{2008ISR}
McCullough, P.\ 2008, STScI Instrument Science Report: 2008-26
\bibitem[Mo et al.(2016)]{2016ApJ...818L..25M} Mo, W., Gonzalez, A., Jee, M.~J., et al.\ 2016, \apjl, 818, L25 
\bibitem[Muzzin et al.(2013)]{2013ApJ...767...39M} Muzzin, A., Wilson, G., Demarco, R., et al.\ 2013, \apj, 767, 39 
\bibitem[Navarro et al.(1997)]{1997ApJ...490..493N} Navarro, J.~F., Frenk, 
C.~S., \& White, S.~D.~M.\ 1997, \apj, 490, 493 
\bibitem[Oguri \& Hamana(2011)]{2011MNRAS.414.1851O} Oguri, M., \& Hamana, T.\ 2011, \mnras, 414, 1851 
\bibitem[Okabe et al.(2010)]{2010PASJ...62..811O} Okabe, N., Takada, M., Umetsu, K., Futamase, T., \& Smith, G.~P.\ 2010, \pasj, 62, 811 
\bibitem[Planck Collaboration et al.(2016)]{2016A&A...594A..13P} Planck Collaboration, Ade, P.~A.~R., Aghanim, N., et al.\ 2016, \aap, 594, A13 
\bibitem[Rafelski et al.(2015)]{2015AJ....150...31R} Rafelski, M., Teplitz, H.~I., Gardner, J.~P., et al.\ 2015, \aj, 150, 31 
\bibitem[Reichardt et al.(2013)]{2013ApJ...763..127R} Reichardt, C.~L., Stalder, B., Bleem, L.~E., et al.\ 2013, \apj, 763, 127 
\bibitem[Rosati et al.(2009)]{2009A&A...508..583R} Rosati, P., Tozzi, P., Gobat, R., et al.\ 2009, \aap, 508, 583 
\bibitem[Rubin et al.(2017)]{2017arXiv170704606R} Rubin, D., Hayden, B., Huang, X., et al.\ 2017, arXiv:1707.04606 
\bibitem[Seitz \& Schneider(1997)]{ss97} Seitz, C.~\& Schneider, P.\ 1997, \aap, 318, 687
\bibitem[Stalder et al.(2013)]{2013ApJ...763...93S} Stalder, B., Ruel, J., {\v S}uhada, R., et al.\ 2013, \apj, 763, 93 
\bibitem[Stanford et al.(2012)]{2012ApJ...753..164S} Stanford, S.~A., Brodwin, M., Gonzalez, A.~H., et al.\ 2012, \apj, 753, 164 
\bibitem[Suzuki et al.(2012)]{2012ApJ...746...85S} Suzuki, N., Rubin, D., Lidman, C., et al.\ 2012, \apj, 746, 85 
\bibitem[Tinker et al.(2008)]{2008ApJ...688..709T} Tinker, J., Kravtsov, A.~V., Klypin, A., et al.\ 2008, \apj, 688, 709-728 
\bibitem[Tozzi et al.(2015)]{2015ApJ...799...93T} Tozzi, P., Santos, J.~S., Jee, M.~J., et al.\ 2015, \apj, 799, 93 
\bibitem[Waizmann et al.(2011)]{2011MNRAS.418..456W} Waizmann, J.-C., Ettori, S., \& Moscardini, L.\ 2011, \mnras, 418, 456 
\bibitem[Wang et al.(2016)]{2016ApJ...828...56W} Wang, T., Elbaz, D., Daddi, E., et al.\ 2016, \apj, 828, 56 
\bibitem[Williamson et al.(2011)]{2011ApJ...738..139W} Williamson, R., Benson, B.~A., High, F.~W., et al.\ 2011, \apj, 738, 139 
\bibitem[Ziparo et al.(2016)]{2016MNRAS.463.4004Z} Ziparo, F., Smith, G.~P., Okabe, N., et al.\ 2016, \mnras, 463, 4004 



\end{thebibliography}
\end{document}